\documentclass[preprint,12pt]{elsarticle}

\usepackage{amssymb}
\usepackage{xspace}
\usepackage{url}
\usepackage{color}
\definecolor{darkred}{rgb}{0.44,0,0}
\definecolor{darkgreen}{rgb}{0,0.44,0}
\definecolor{darkblue}{rgb}{0,0,0.44}
\definecolor{grey}{rgb}{0.5,0.5,0.5}
\usepackage{multirow}

\journal{Journal of Computational Science}

\begin{document}

\newcommand{\gemm}{{\sc gemm}\xspace}
\newcommand{\gepp}{{\sc gepp}\xspace}
\newcommand{\gemp}{{\sc gemp}\xspace}
\newcommand{\gepm}{{\sc gepm}\xspace}
\newcommand{\symm}{{\sc symm}\xspace}
\newcommand{\hemm}{{\sc hemm}\xspace}
\newcommand{\sypm}{{\sc sypm}\xspace}
\newcommand{\symp}{{\sc symp}\xspace}
\newcommand{\trmm}{{\sc trmm}\xspace}
\newcommand{\trpm}{{\sc trpm}\xspace}
\newcommand{\trmp}{{\sc trmp}\xspace}
\newcommand{\trsm}{{\sc trsm}\xspace}
\newcommand{\trps}{{\sc trps}\xspace}
\newcommand{\trsp}{{\sc trsp}\xspace}
\newcommand{\syrk}{{\sc syrk}\xspace}
\newcommand{\syrtk}{{\sc syr2k}\xspace}
\newcommand{\herk}{{\sc herk}\xspace}
\newcommand{\hertk}{{\sc her2k}\xspace}

\newcommand{\geqrf}{{\sc geqrf}\xspace}
\newcommand{\getrf}{{\sc getrf}\xspace}
\newcommand{\potrf}{{\sc potrf}\xspace}
\newcommand{\sytrd}{{\sc sytrd}\xspace}
\newcommand{\symv}{{\sc symv}\xspace}

\begin{frontmatter}



\title{Multi-Threaded Dense Linear Algebra Libraries for\\
       Low-Power Asymmetric Multicore Processors}
       
\author[uji]{Sandra~Catal\'an}
\ead{catalans@uji.es}
\author[upc]{Jos\'e~R.~Herrero}
\ead{josepr@ac.upc.edu}
\author[ucm]{Francisco~D.~Igual}
\ead{figual@ucm.es}
\author[uji]{Rafael~Rodr\'{\i}guez-S\'anchez}
\ead{rarodrig@uji.es}
\author[uji]{Enrique~S.~Quintana-Ort\'{\i}}
\ead{quintana@uji.es}

\address[uji]{Depto. Ingenier\'{\i}a y Ciencia de Computadores,
             Universidad Jaume I, Castell\'on, Spain.}
\address[upc]{Dept. d'Arquitectura de Computadors, Universitat Polit\`ecnica de Catalunya, Spain.}       
\address[ucm]{Depto. de Arquitectura de Computadores y Autom\'atica, Universidad Complutense de Madrid, Spain.}       


\author{}

\address{}

\begin{abstract}
Dense linear algebra libraries, such as BLAS and LAPACK, provide a relevant collection of numerical tools 
for many scientific and engineering applications.
While there exist high performance implementations of the BLAS (and LAPACK) functionality
for many current multi-threaded architectures, the adaption of these libraries for
asymmetric multicore processors (AMPs) is still pending. In this paper we address this challenge
by developing an asymmetry-aware implementation of the BLAS, based on the
BLIS framework, and tailored for AMPs equipped with two types of cores: fast/power hungry versus slow/energy efficient.
For this purpose, we integrate coarse-grain and fine-grain parallelization strategies into the library routines
which, respectively, dynamically distribute the workload between the two core types and statically  
repartition this work among the cores of the same type.

Our results on an  ARM\textregistered\ big.LITTLE\texttrademark\ processor embedded in the Exynos 5422 SoC,
using the asymmetry-aware version of the BLAS and a plain migration of the legacy version of LAPACK,
experimentally assess the benefits, limitations, and potential of this approach.

\end{abstract}


\end{frontmatter}


\section{Introduction}

Dense linear algebra (DLA) is at the bottom of the ``food chain'' for many scientific and engineering applications, which can be often
decomposed into a collection of 
linear systems of equations, linear least squares (LLS) problems, rank-revealing computations, and
eigenvalue problems~\cite{dem97}.
The importance of these linear algebra operations is well recognized and, from the numerical point of view, 
when they involve {\em dense} matrices,
their solution can be reliably addressed using the {\em Linear Algebra PACKage} (LAPACK)~\cite{lapack}. 

To attain portable performance,
LAPACK routines cast a major fraction of their computations in terms of a reduced number of {\em Basic Linear Algebra Subprograms} (BLAS)~\cite{blas1,blas2,blas3},
employing an implementation of the BLAS specifically optimized for the target platform.
Therefore, it comes as no surprise that
nowadays there exist both commercial and open source implementations of the BLAS targeting 
a plethora of architectures, available among others in
AMD ACML~\cite{acml}, IBM ESSL~\cite{essl}, Intel MKL~\cite{mkl}, NVIDIA CUBLAS~\cite{cublas}, 
ATLAS~\cite{atlas}, GotoBLAS~\cite{Goto:2008:AHP}, OpenBLAS~\cite{OpenBLAS}, and BLIS~\cite{BLIS1}.
Many of these implementations offer multi-threaded kernels that can
exploit the hardware parallelism of a general-purpose multicore processor or, in the case of NVIDIA's BLAS, 
even those in a many-core graphics processing unit (GPU). 

Asymmetric multicore processors (AMPs), such as the recent ARM\textregistered\ big.LITTLE\texttrademark\ systems-on-chip (SoC),
are a particular class of heterogeneous architectures
that combine 
a few high performance (but power hungry) cores with a collection of energy efficient (though slower) cores.\footnote{%
AMPs differ from a heterogeneous SoC like the NVIDIA Tegra TK1, in that the cores of the AMP share the same instruction set architecture (ISA).}
With the end of 
Dennard scaling~\cite{Den74}, but
the steady doubling of transistors in CMOS chips at the pace 
dictated by Moore's law~\cite{Moo65}, 
AMPs have gained considerable appeal as, in theory,
they can deliver much higher performance for the same power budget~\cite{Kum04,Hil08,Mor06,Win10}.

In past work~\cite{asymBLIS}, we demonstrated how to adapt BLIS in order to attain high performance 
for the multiplication of two square matrices,
on an ARM big.LITTLE AMP consisting of ARM Cortex-A15 and Cortex-A7 clusters. 
In this paper, we significantly extend our previous work by applying similar parallelization principles to
the complete Level-3 BLAS (BLAS-3), 
and we evaluate the impact of these optimizations on LAPACK. 
In particular, our work makes the following contributions:
\begin{itemize}
\item Starting from the reference implementation of the BLIS library (version 0.1.8), 
      we develop a multi-threaded parallelization of the complete BLAS-3 for
      any generic AMPs, tailoring it for the ARM big.LITTLE AMP embedded in the Exynos~5422 SoC in particular.
      These tuned kernels not only distinguish between different operations (e.g., paying special care
      to the parallelization of the triangular system solve), but also take into consideration the operands' dimensions (shapes).
      This is especially interesting because, in general, the BLAS-3 are often invoked from LAPACK to operate on highly non-square 
      matrix blocks.
\item We validate the correction of the new BLIS-3 by integrating them with the legacy implementation of
      LAPACK (version 3.5.0) from the netlib public repository.\footnote{Available at \url{http://www.netlib.org/lapack}.}
\item We illustrate the practical performance that can be attained from a straight-forward migration and
      execution of LAPACK, on top of the new BLIS-3 for the Exynos~5422, that basically adjusts the algorithmic
      block sizes and only carries out other minor modifications.\\
      In particular, our experiments with three relevant matrix routines from LAPACK, key for the solution of linear systems
      and symmetric eigenvalue problems, show a case of success for a matrix factorization routine; 
      a second scenario where a significant modification of the LAPACK
      routine could yield important performance gains; and a third case where performance is limited by the
      memory bandwidth, but a multi-threaded implementation of the Level-2 BLAS~\cite{blas2} could render a moderate
      improvement in the results. 
\end{itemize}
To conclude, we emphasize that the general parallelization approach proposed in this paper for AMPs
can be ported, with little effort, to present and future instances of the ARM big.LITTLE architecture
as well as to any other asymmetric design in general (e.g. the Intel QuickIA prototype~\cite{IntelQuickIA},
or general-purpose SMPs with cores running at different frequencies).

The rest of the paper is structured as follows.
In Section~\ref{sec:related}, we briefly review the foundations of BLIS, and we discuss two distinct
approaches (though complementary under certain conditions) to extract
parallelism from LAPACK, based on a runtime that exploits task-parallelism and/or by leveraging a multi-threaded
implementation of the BLAS. 
In Section~\ref{sec:blis}, we introduce and evaluate our multi-threaded implementation of the complete BLIS-3, for
matrix operands of distinct shapes, tuned for the big.LITTLE AMP architecture in the Exynos 5422 SoC. 
In Section~\ref{sec:lapack}, we illustrate the impact of leveraging our platform-specific BLIS-3 from LAPACK using
three key operations.
Finally, in Section~\ref{sec:remarks} we offer a few concluding remarks and discuss future work.

\section{BLIS and other Related Work}
\label{sec:related}

\subsection{BLIS}

The conventional and easiest
approach to obtain a parallel execution of LAPACK, on a multicore architecture, simply leverages 
a multi-threaded implementation of the BLAS that partitions the work among the computational resources, 
thus isolating LAPACK from this task.
For problems of small to moderate dimension, platforms with a low number of cores, and/or
DLA operations with simple data dependencies (like those in the BLAS-3), 
this approach usually provides optimal efficiency.
Indeed, this is basically the preferred option adopted by 
many commercial implementations of LAPACK.

Most modern implementations of the BLAS follow the path pioneered by 
GotoBLAS to implement the kernels in BLAS-3 
as three nested loops around two packing routines, 
which orchestrate the transfer of data between consecutive levels of the cache-memory hierarchy, and a macro-kernel in charge of performing the actual computations.
BLIS internally decomposes the macro-kernel into two additional loops around
a micro-kernel that, in turn, is implemented as a loop around a symmetric rank-1 update (see Figure~\ref{fig:gotoblas_gemm}).
In practice, the micro-kernel is encoded in assembly or in C enhanced with vector intrinsics; see~\cite{BLIS1} for details. 

\begin{figure}[t]
\centering
\begin{minipage}[c]{\textwidth}
\footnotesize
\resizebox{\linewidth}{!}{
\begin{tabular}{llll}
Loop 1 &{\bf for} $j_c$ = $0,\ldots,n-1$ {\bf in steps of} $n_c$\\
Loop 2 & \hspace{3ex}  {\bf for} $p_c$ = $0,\ldots,k-1$ {\bf in steps of} $k_c$\\
&\hspace{6ex}           \textcolor{darkblue}{$B(p_c:p_c+k_c-1,j_c:j_c+n_c-1)$} $\rightarrow \textcolor{darkblue}{B_c}$ & & // Pack into $B_c$\\
Loop 3 & \hspace{6ex}           {\bf for} $i_c$ = $0,\ldots,m-1$ {\bf in steps of} $m_c$\\
&\hspace{9ex}                     \textcolor{darkred}{$A(i_c:i_c+m_c-1,p_c:p_c+k_c-1)$} $\rightarrow \textcolor{darkred}{A_c}$ & & // Pack into $A_c$ \\
\cline{2-4}
Loop 4&\hspace{9ex} {\bf for} $j_r$ = $0,\ldots,n_c-1$ {\bf in steps of} $n_r$  & & // Macro-kernel\\
Loop 5&\hspace{12ex}   {\bf for} $i_r$ = $0,\ldots,m_c-1$ {\bf in steps of} $m_r$\\
\cline{2-3}
&\hspace{15ex}             \textcolor{darkgreen}{$C_c(i_r:i_r+m_r-1,j_r:j_r+n_r-1)$} & // Micro-kernel \\
&\hspace{19ex} ~$\mathrel{+}=$     ~\textcolor{darkred}{$A_c(i_r:i_r+m_r-1,0:k_c-1)$} \\
&\hspace{19ex} ~~~$\cdot$~~~~\textcolor{darkblue}{$B_c(0:k_c-1,j_r:j_r+n_r-1)$} \\
\cline{2-3}
&\hspace{12ex} {\bf endfor}\\
&\hspace{9ex} {\bf endfor}\\
\cline{2-4}
&\hspace{6ex} {\bf endfor}\\
&\hspace{3ex} {\bf endfor}\\
&{\bf endfor}\\
\end{tabular}
}
\end{minipage}
\caption{High performance implementation of the matrix multiplication in BLIS. In the code, $C_c \equiv C(i_c:i_c+m_c-1,j_c:j_c+n_c-1)$
is just a notation artifact, introduced to ease the presentation of the algorithm, while $A_c,B_c$ correspond to actual buffers that are involved in data copies.}
\label{fig:gotoblas_gemm}
\end{figure}

A multi-threaded parallelization of the matrix multiplication (\gemm) in BLIS for
conventional symmetric multicore processors (SMPs)
and modern many-threaded architectures was presented in~\cite{BLIS2,BLIS3}.
These parallel implementations exploit
the concurrency available in
the nested five--loop organization of \gemm at one or more levels
(i.e., loops), taking into account the cache organization of the target platform,
the granularity of the computations, and the risk of race conditions, among other factors.

In~\cite{asymBLIS} we leverage similar design principles to
propose a high performance implementation of the \gemm kernel from BLIS for an ARM big.LITTLE SoC with two quad-core clusters, 
consisting of ARM Cortex-A15 and ARM Cortex-A7 cores. 
Specifically, starting from the BLIS code for \gemm, we modify the loop stride configuration and scheduling policy
to carefully distribute the micro-kernels comprised by certain loops among the ARM Cortex-A15 and Cortex-A7 
clusters and cores taking into consideration
their computational power and cache organization.

\subsection{Runtime-based task-parallel LAPACK}

Extracting task parallelism has been recently proved to yield an efficient means to tackle the computational power of current
general-purpose multicore and many-core processors.
Following the path pioneered by Cilk~\cite{cilkweb}, 
several research efforts ease the development and improve 
the performance of task-parallel programs by embedding task scheduling inside a
{\em runtime}. The benefits of this approach for complex DLA operations
have been reported, among others, by OmpSs~\cite{ompssweb}, StarPU~\cite{starpuweb}, PLASMA~\cite{plasmaweb,magmaweb},
Kaapi~\cite{kaapiweb}, and {\tt libflame}~\cite{flameweb}.
In general, the runtimes underlying these tools decompose a DLA routine into a collection
of numerical kernels (or {\em tasks}),
and then take into account the dependencies between the tasks in order to correctly issue their execution to the system cores.
The tasks are therefore the ``indivisible'' scheduling unit while the cores constitute the basic computational resources.

The application of a runtime-based approach to schedule DLA operations in an AMP is still
quite immature.
Botlev-OmpSs~\cite{OmpSsbigLITTLE} is an instance of the OmpSs runtime that 
embeds a Criticality-Aware Task Scheduler (CATS)
specifically designed with AMPs in mind.
This asymmetry-conscious runtime relies on bottom-level longest-path priorities,
and keeps track of the criticality of the individual tasks to
place them in either a critical queue or a non-critical one.
Furthermore, tasks enqueued in the critical queue can only be executed by the fast cores,
and the enhanced scheduler integrates uni- or bi-directional work stealing between fast and slow cores.

Botlev-OmpSs required an important redesign of the underlying scheduling policy to exploit the asymmetric architecture.
Alternatively, in~\cite{asymChol} we proposed an approach
to refactor any asymmetry-oblivious runtime task scheduler by 
{\em i)} aggregating the cores of the AMP into a number of {\em symmetric virtual cores},
    which become the only computational resources visible to the runtime scheduler; and
{\em ii)} hiding the difficulties intrinsic to dealing with an asymmetric architecture 
    inside an asymmetry-aware implementation of the BLAS-3.

The benefits of these two AMP-specific approaches have been demonstrated in~\cite{OmpSsbigLITTLE,asymChol}
for the Cholesky factorization. Unfortunately, applying the same principles to the full contents of a library as
complex as LAPACK is a daunting task. First, one would need to transform all the algorithms underlying the library
to produce task-parallel versions, which can then be adapted for and feed to a specific runtime scheduler. While this work has been
done for some combinations of basic matrix factorizations
(for the solution of linear systems, LLS, and eigenvalue problems), runtimes, 
and target platforms~\cite{BadiaHLPQQ09,Buttari200938,Quintana:2008:PMA}, the effort is far from negligible.

In this paper we depart from previous work by hiding the asymmetry-aware optimization inside
a parallel implementation of the complete BLAS-3 for ARM big.LITTLE architectures, which is then
invoked from the legacy implementation of LAPACK. We note that, though we do not address 
the BLAS-1 and BLAS-2 in our work, the parallelization of the kernels in these two levels of the BLAS 
is straight-forward, even for an AMP.

\section{Asymmetric-Aware BLAS for the Exynos 5422 SoC}
\label{sec:blis}

\subsection{Target architecture}

\begin{figure}[t]
\begin{center}
\includegraphics[width=0.6\columnwidth]{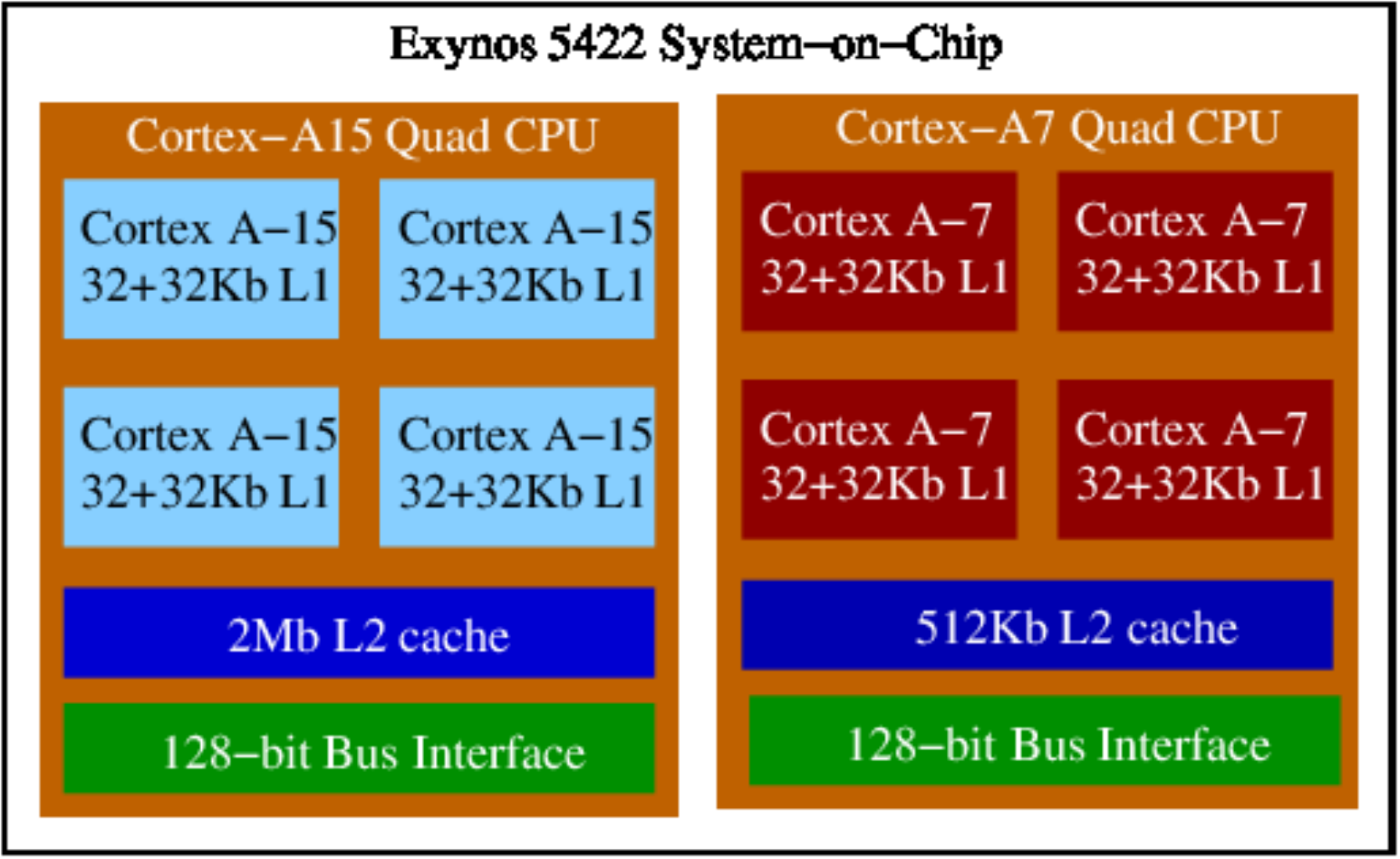}
\end{center}
\caption{ Exynos 5422 block diagram.}
\label{fig:exynos}
\end{figure}

The AMP employed in the experimentation is an ODROID-XU3 board furnished with a Samsung Exynos 5422 SoC. This processor 
comprises an ARM
Cortex-A15 quad-core processing cluster (1.4~GHz) plus a Cortex-A7 quad-core processing
cluster (1.6~GHz). Each Cortex core has its own private 32+32-Kbyte L1 (instruction+data) cache.
The four ARM Cortex-A15 cores share a 2-Mbyte L2 cache, and the four ARM Cortex-A7 cores share a smaller 512-Kbyte L2 cache.
In addition, the two clusters access a DDR3 RAM (2~Gbytes) via 128-bit coherent bus interfaces; see Figure~\ref{fig:exynos}.

\subsection{BLIS kernels}

The specification of the BLAS-3~\cite{blas3} basically comprises 6--9~kernels offering the following functionality: 
\begin{enumerate}
\itemsep=0pt\parskip=0pt
\item Compute (general) matrix multiplication (\gemm), 
      as well as specialized versions of this operation where one of the input operands is 
      symmetric/Hermitian (\symm/\hemm) or triangular (\trmm).
\item Solve a triangular linear system (\trsm).
\item Compute a symmetric/Hermitian rank-$k$ or rank-$2k$ update 
      (\syrk/\herk or \syrtk/\hertk, respectively).
\end{enumerate}
The specification accommodates two data types (real or complex) and
two precisions (single or double), as well as operands with different
``properties'' (e.g., upper/lower triangular, transpose or not, etc.).
Note that \hemm, \herk, and \hertk are only defined for complex data, providing the same functionality as
\symm, \syrk, and \syrtk for real data.

For brevity, in the following study we will address the real double-precision version of these operations. 
Furthermore, we will only target the cases in Table~\ref{tab:blis-3},
where we will consider upper triangular matrices and we will operate with/on the upper
triangular part of symmetric matrices.
However, we note that, due to the organization of BLIS, our optimized implementations 
for the Exynos 5422 SoC cover all other cases.

\begin{table}[th!]
{\footnotesize
\begin{center}
\begin{tabular}{|l||l|c|c|c|}
\hline
Kernel                      & Operation         & \multicolumn{3}{c|}{Operands} \\ 
                            &                   & $A$ & $B$ &$C$                \\ \hline \hline
\gemm                       & $C:=C+AB$         & $m \times k$           & $k \times n$ & $m \times n$ \\  \hline \hline
\multirow{2}{*}{\symm}      & $C:=C+AB$ or      & Symmetric $m\times m$  & \multirow{2}{*}{$m \times n$} & \multirow{2}{*}{$m \times n$} \\
                            & $C:=C+BA$         & Symmetric $n\times n$  & & \\ \hline
\multirow{2}{*}{\trmm}      & $B:=AB$ or        & Triangular $m\times m$ & \multirow{2}{*}{$m \times n$} & \multirow{2}{*}{--} \\
                            & $B:=BA$           & Triangular $n\times n$ & & \\ \hline \hline
\multirow{2}{*}{\trsm}      & $B:=A^{-1}B$ or   & Triangular $m\times m$ & \multirow{2}{*}{$m \times n$} & \multirow{2}{*}{--} \\
                            & $B:=BA^{-1}$      & Triangular $n\times n$ & & \\ \hline\hline
\syrk                       & $C:=C+A^TA$       & $k \times n$           & -- & $n \times n$ \\  \hline 
\syrtk                      & $C:=C+A^TB+B^TA$  & $k \times n$           & $k\times n$ & $n \times n$ \\  \hline 
\end{tabular}
\end{center}
}
\caption{Kernels of BLIS-3 considered in the evaluation.}
\label{tab:blis-3}
\end{table}

The steps to attain high performance from these kernels in the Exynos 5422 SoC require:
\begin{enumerate}
\itemsep=0pt\parskip=0pt
\item 
       to develop highly optimized implementations of the underlying micro-kernels for the ARM Cortex-A15 and Cortex-A7 cores;
\item 
       to tune the configuration parameters $m_c,n_c,k_c$ to the target type of core; and
\item 
       to enforce a balanced distribution of the workload between both types of cores.
\end{enumerate}
The following subsection offers some hints on the first two tasks, which have been carried out following a
development and experimental optimization approach similar to those necessary in a homogeneous (non-asymmetric) architecture.

Our major contribution is introduced next, in
subsection~\ref{subsec:dynamic}, where we investigate the best parallelization strategy 
depending on the kernel and the operands' shape.
This is a crucial task as, in practice, 
the invocations to the BLAS-3 kernels from LAPACK generally involve nonsquare operands with one (or more)
small dimension(s).

\subsection{Cache optimization of BLIS}

For the ARM Cortex-A15 and Cortex-A7 core architectures, the BLIS micro-kernels are manually encoded with $m_r=n_r=4$. Furthermore, via an extensive
experimental study, the configuration parameters are set to
$m_c = (152,80)$ for the ARM (Cortex-A15,Cortex-A7) cores;
and $k_c = 352, n_c = 4096$ for both types of cores. 
With these values, 
the buffer $A_c$, of dimension $m_c \times k_c$ (408~KB for the Cortex-A15 and 215~KB for the Cortex-A7), 
fits into the L2 cache of the corresponding cluster,
while a micro-panel of $B_c$, of dimension $k_c \times n_r$ (11~KB), fits into the L1 cache of the each core.
The micro-kernel thus streams $A_c$ together with the micro-panel of $B_c$ into the 
floating-point units (FPUs) from the L2 and L1 caches, respectively; see~\cite{asymBLIS} for details.

\subsection{Multi-threaded parallelization of the BLIS-3}
\label{subsec:dynamic}

\subsubsection{Square operands in \gemm}

The {\em asymmetry-aware parallelization} of this kernel in~\cite{asymBLIS} 
targeted only a matrix multiplication with square operands ($m=n=k$), applying
a {\em dynamic schedule} to Loop~3
in order to distribute its iteration space between the two types of clusters (coarse-grain partitioning).
In addition, a {\em static schedule} was internally applied to distribute
the iteration space of either Loop~4 or Loop~5 among the cores of the same cluster (fine-grain partitioning).
The parallelization of Loop~1 was discarded because $n_c=4096$, and this large value turns very difficult
to attain a balanced workload distribution between the two clusters. 
The parallelization of Loops~2 and/or~6 was
also discarded, because they both
require a synchronization mechanism to deal with race conditions.

Following the solution adopted in~\cite{asymBLIS}, we will 
use ``{\tt D3S4}'' and ``{\tt D3S5}''  to refer to strategies based on
a dynamic coarse-grain parallelization of
Loop~3 combined with a static fine-grain parallelization of either Loop~4 or Loop~5, respectively.
To assess the efficiency of these two options, we will measure the GFLOPS rates (billions of floating-point
arithmetic operations per second) they attain and compare those against an ``{\tt ideal}'' execution
where the eight cores incur no access conflicts and the workload is perfectly balanced.
To estimate the latter, we will experimentally evaluate the GFLOPS achieved with the serial BLIS kernel, 
using either a single ARM Cortex-A15 core or a single ARM Cortex-A7 core, 
and then consider the ideal peak performance as the aggregation of both rates multiplied by 4. 

Unless otherwise stated,
the stride configuration parameter for Loop~3 is set to $m_c = (152,32)$ for the
ARM (Cortex-A15,Cortex-A7) cores. 
The value selected for the Cortex-A7 architecture is thus smaller than the experimental optimal ($m_c=80$), but 
this compromise was adopted to roughly match the ratio between the computational power of both types of cores as well as to improve
the workload distribution.

The top-left plot in
Figure~\ref{fig:gemm} reports the performance 
attained with the dynamic-static parallelization strategies 
for a matrix multiplication involving square operands only.
The results show that the two options, {\tt D3S4} and {\tt D3S5}, 
obtain a large fraction of the GFLOPS rate estimated for the ideal scenario,
though the combination that parallelizes Loops~3+4 is consistently better.
Concretely, from $m=n=k\geq 2000$, this option delivers between 12.4 and 12.7~GFLOPS,
which roughly represents 93\% of the ideal peak performance. 
In this plot, we also include the results for an strategy that parallelizes 
Loop~4 only, distributing its workload among the ARM Cortex-A15 and Cortex-A7 cores, but oblivious of their different
computational power (line labelled as ``{\tt ObS4}''). With this asymmetry-agnostic option, 
the synchronization at the end of the
parallel regions slows down the ARM Cortex-A15 cores, yielding the poor GFLOPS rate observed
in the plot. 

\subsubsection{\gemm with rectangular operands}

\begin{figure}[th!]
\begin{center}
\begin{tabular}{c}
\begin{minipage}[c]{\textwidth}
\includegraphics[width=0.5\textwidth]{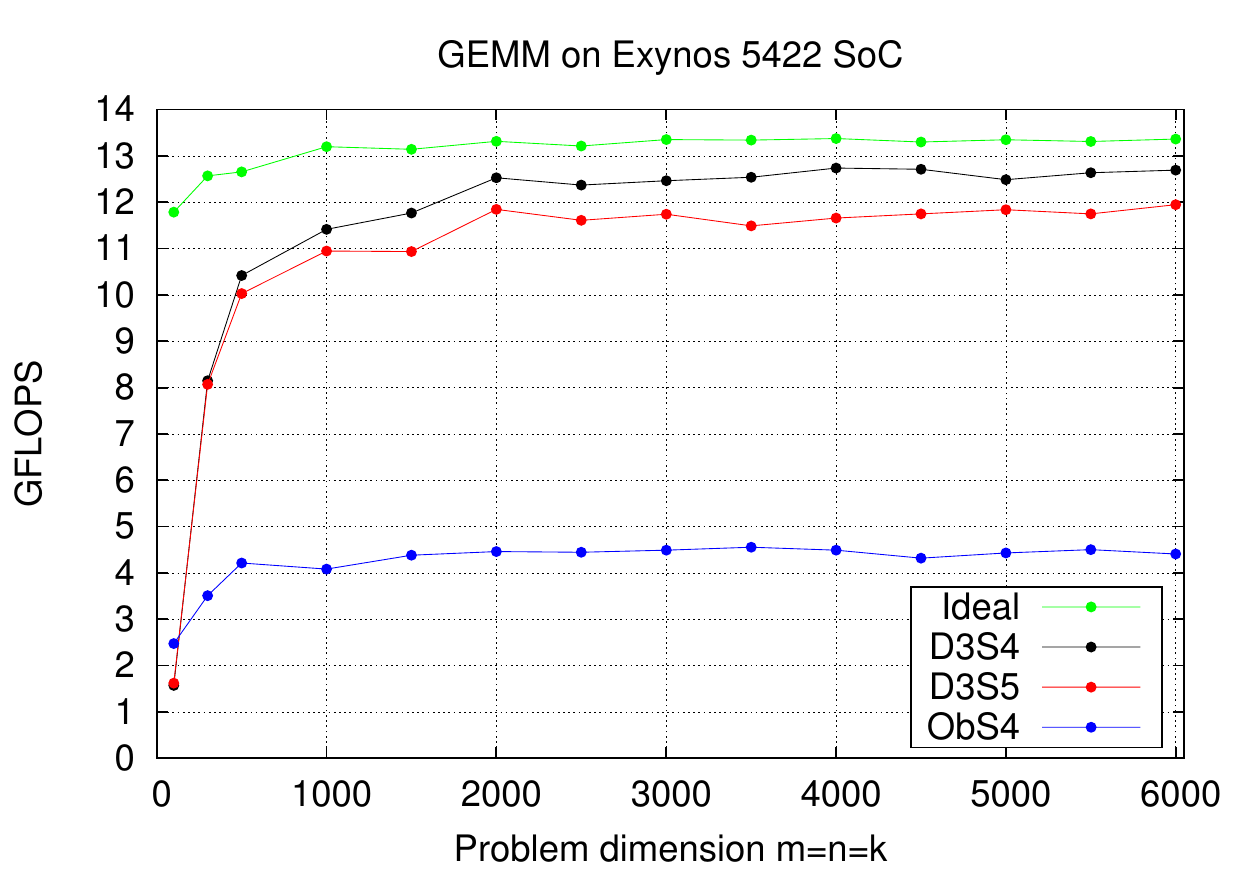}
\includegraphics[width=0.5\textwidth]{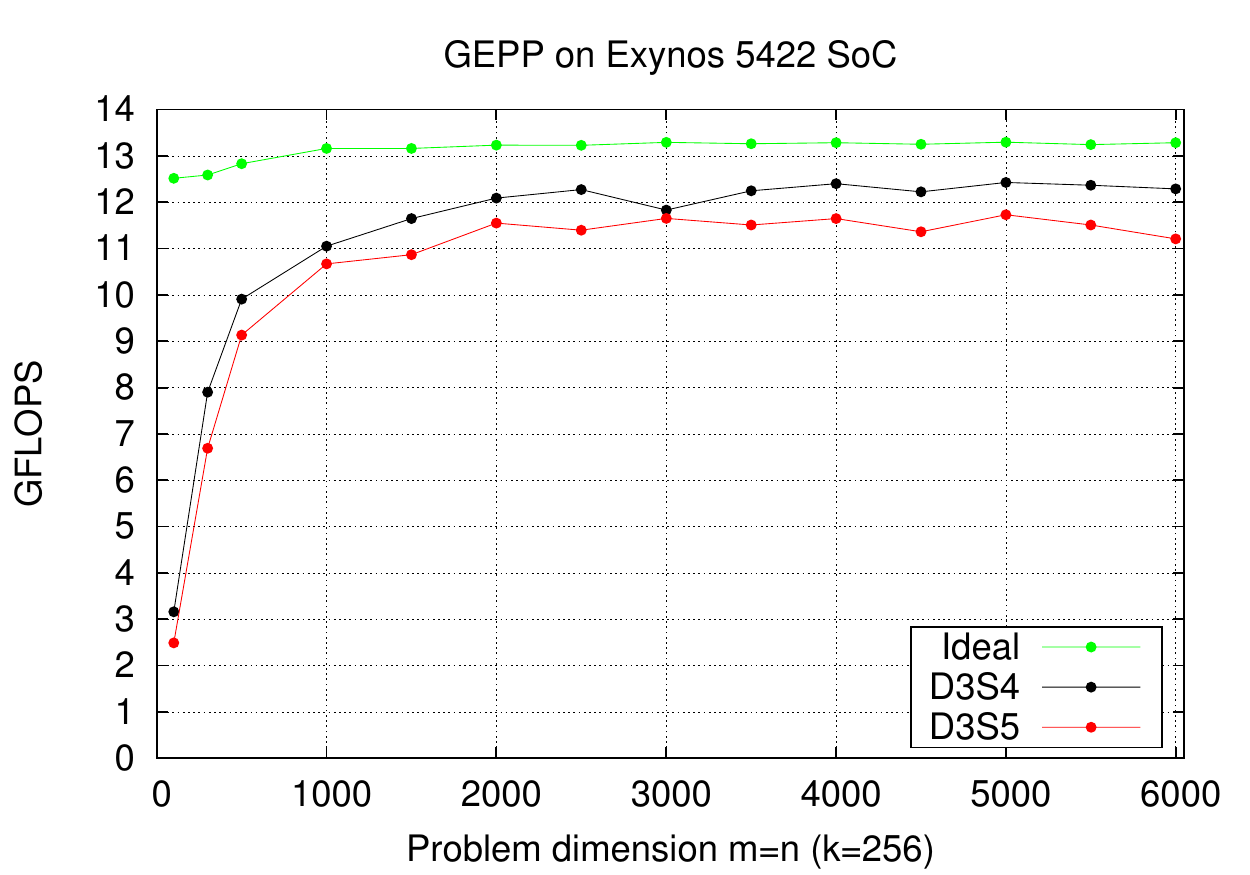}\\
\includegraphics[width=0.5\textwidth]{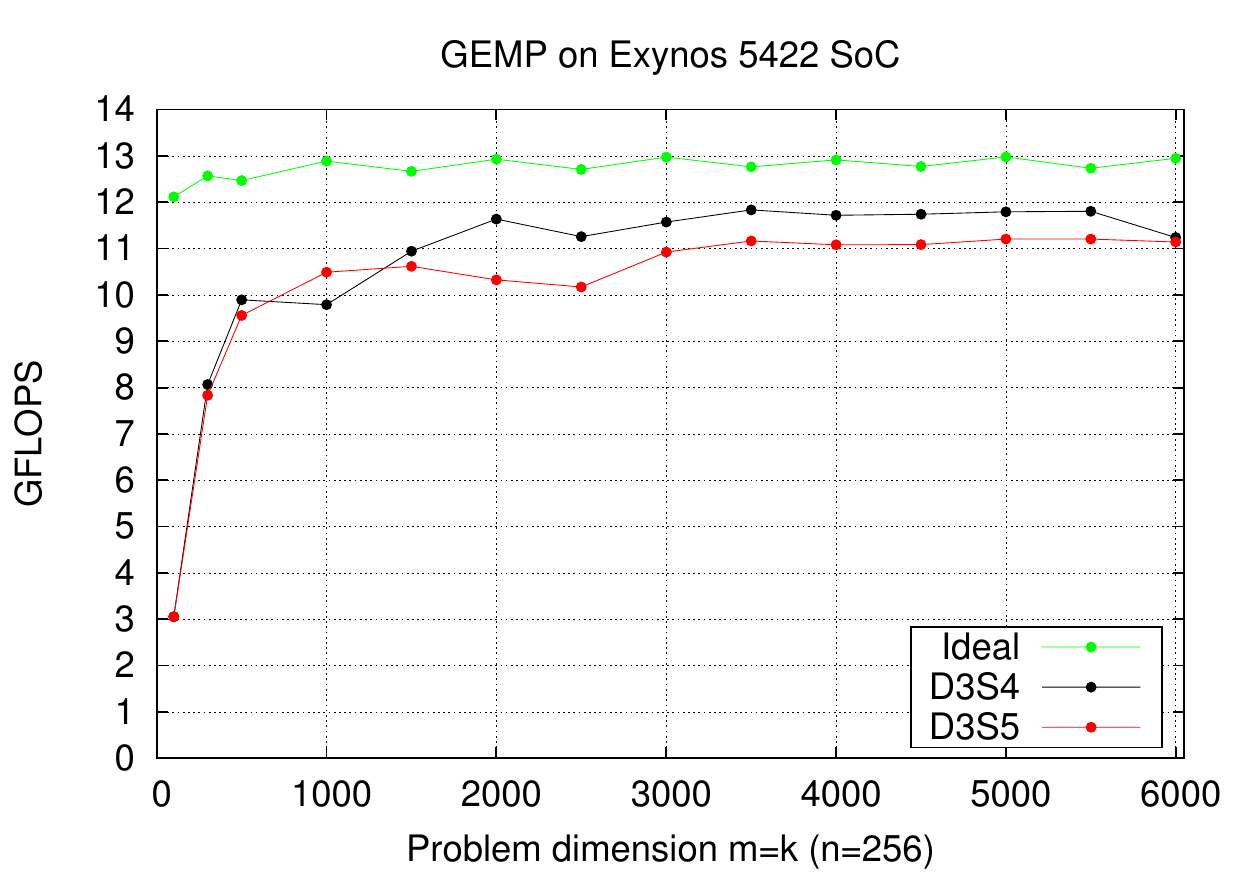}
\includegraphics[width=0.5\textwidth]{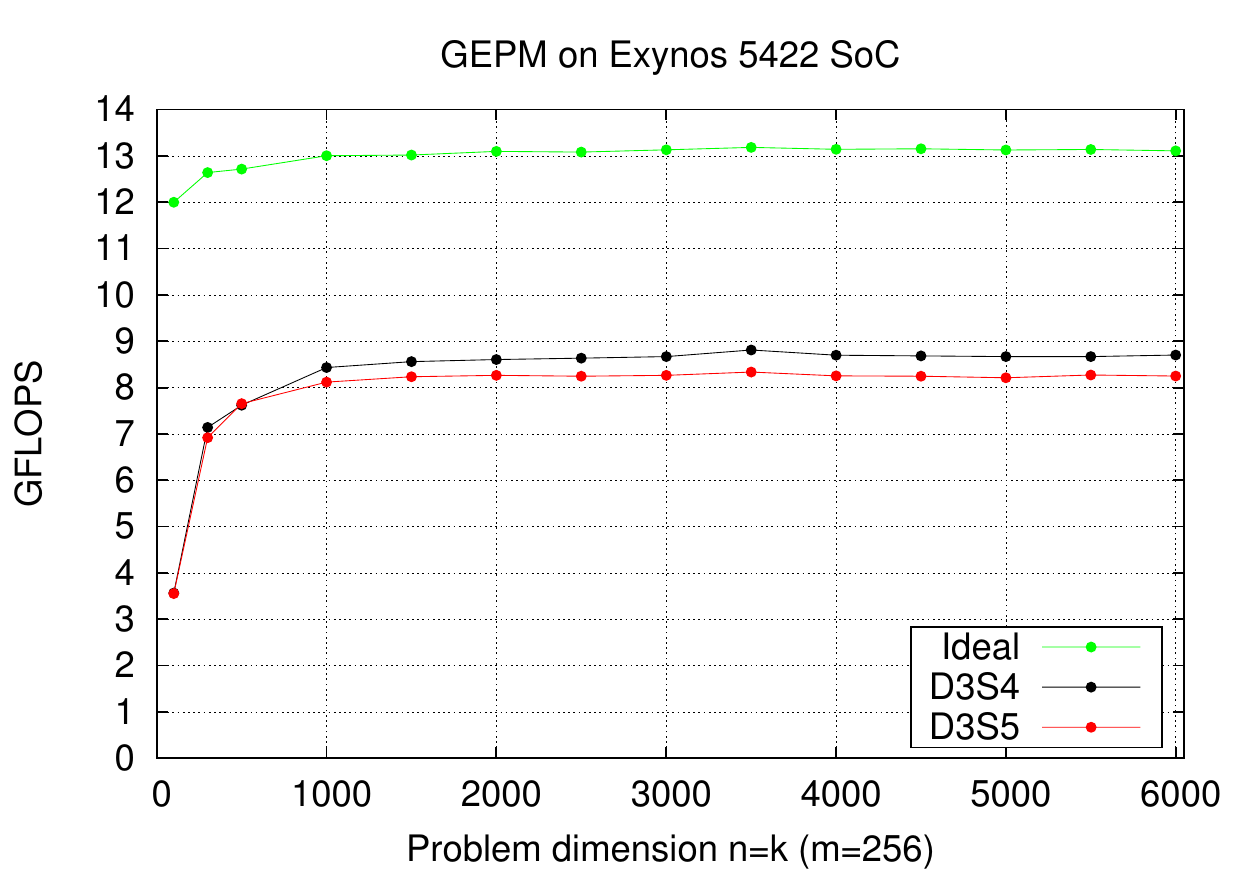}
\end{minipage}
\end{tabular}
\end{center}
\caption{Performance of (general) matrix multiplication with square matrices:
         \gemm; and three rectangular cases with two equal dimensions: 
         \gepp, 
         \gemp, and 
         \gepm.} 
\label{fig:gemm}
\end{figure}

The remaining three plots in Figure~\ref{fig:gemm} report the performance of the asymmetry-aware parallelization strategies 
when the matrix multiplication kernel is invoked, (e.g., from LAPACK,) 
to compute a product 
for the following ``rectangular'' cases (see Table~\ref{tab:blis-3}):
\begin{enumerate}{}{}
\itemsep=0pt\parskip=0pt
\item 
      \gepp (general panel-panel multiplication)  for $m=n\neq k$;
\item 
      \gemp (general matrix-panel multiplication) for $m=k\neq n$; and
\item 
      \gepm (general panel-matrix multiplication) for $n=k\neq m$.
\end{enumerate}
In these three specialized cases, we vary the two equal dimensions in the range ${\cal R}=\{100,300,500,1000,1500,\ldots,6000\}$ 
and fix the remaining one to $256$.
(This specific value was selected because it is often 
used as the algorithmic block size for many LAPACK routines/target architectures.)

The plots for \gepp and \gemp (top-right and bottom-left in Figure~\ref{fig:gemm}) 
show GFLOPS rates that are similar to those attained 
when the same strategies are applied to the ``square case'' (top-left plot in the same figure), 
with {\tt D3S4} outperforming {\tt D3S5} again.
Furthermore, the performances attained with this particular strategy,
when the variable problem dimension is equal or larger than 2000 
(11.8--12.4 GFLOPS for \gepp and 11.2--11.8 GFLOPS for \gemp), 
is around 90\% of those expected in an ideal scenario.
We can thus conclude that, for these particular matrix shapes, this specific parallelization option is reasonable.

\begin{figure}[th!]
\begin{center}
\begin{tabular}{c}
\begin{minipage}[c]{\textwidth}
\includegraphics[width=0.8\textwidth]{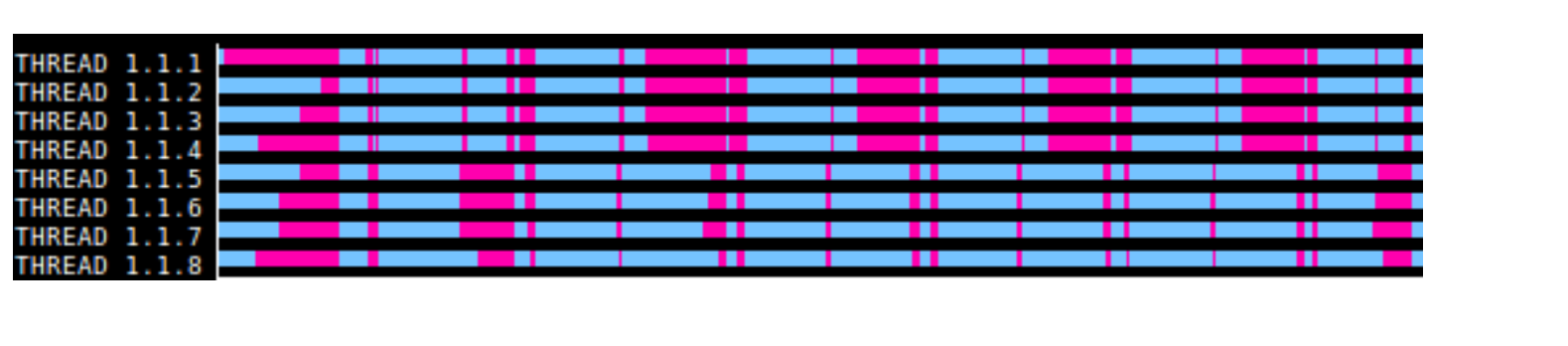}\\
\includegraphics[width=0.8\textwidth]{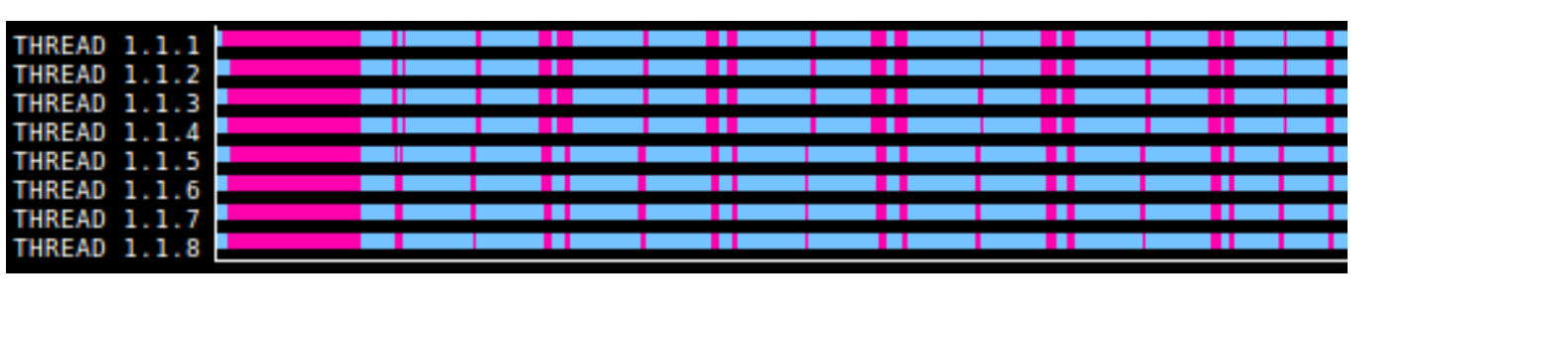}
\
\end{minipage}
\end{tabular}
\end{center}
\caption{Execution traces of \gepm using the parallelization strategy {\tt D3S4} for 
         a problem of dimension $n=k=2000$ and $m=256$. The top plot 
         corresponds to the use of cache configuration parameters $m_c=(152,32)$ for the ARM (Cortex-A15,Cortex-A7) cores,
         respectively. The bottom plot reduces these values to $m_c=(116,24)$. The blue periods correspond to actual work while
         the pink ones represent synchronization (idle time).}
\label{fig:traces}
\end{figure}

The application of the same strategies to \gepm delivers mediocre results, though.
The reason is that, when $m=256$, a coarse-grain distribution of the workload that assigns chunks
of $m_c=(152,32)$ iterations of Loop~3 to the ARM (Cortex-A15,Cortex-A7) cores may be appropriate from the point of view of the
cache utilization, but yields a highly unbalanced execution.
This behaviour is exposed with an execution trace, obtained with the {\tt Extrae} framework~\cite{extrae},
in the top part of Figure~\ref{fig:traces}.

\begin{figure}[th!]
\begin{center}
\includegraphics[width=0.5\textwidth]{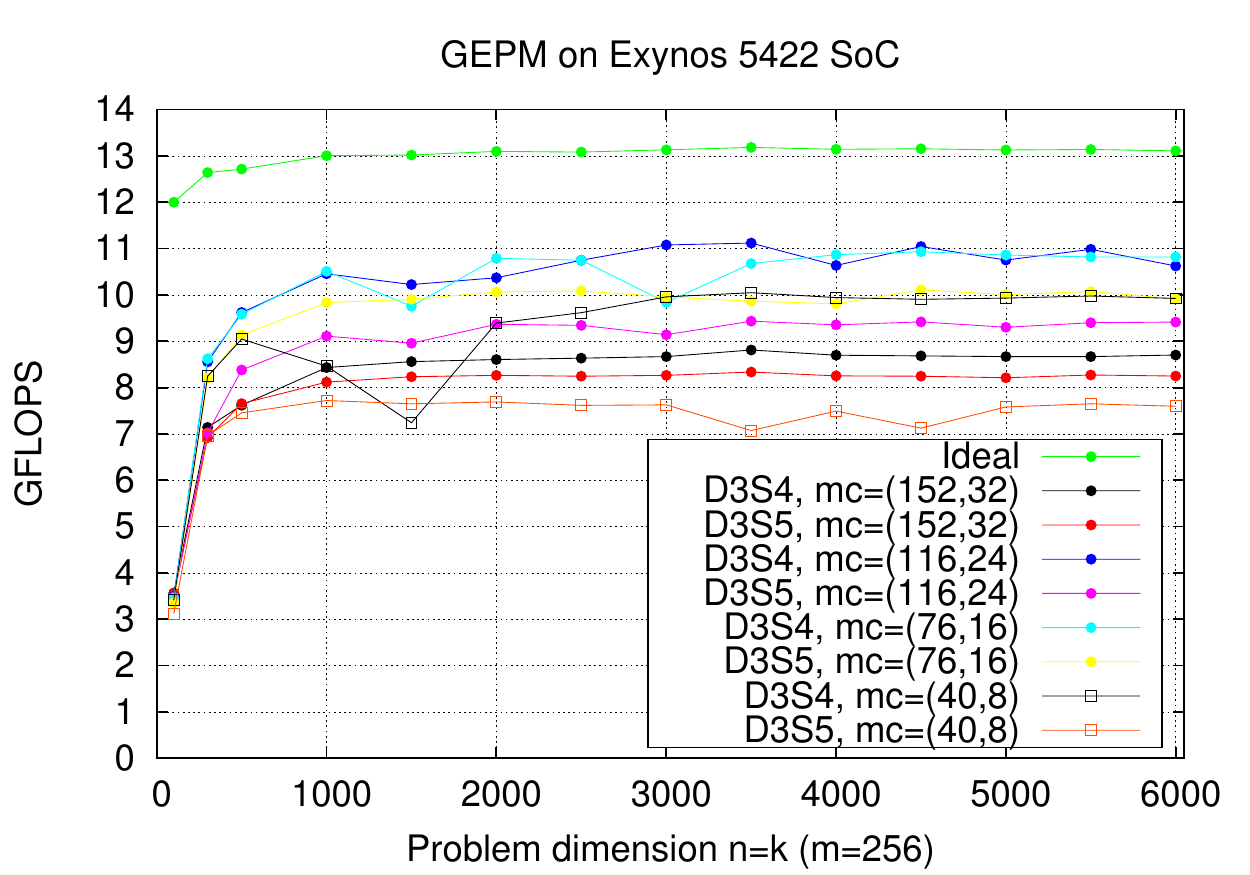}
\end{center}
\caption{Performance of \gepm for different cache configuration parameters $m_c$.}
\label{fig:gepm}
\end{figure}

To tackle the unbalanced workload distribution problem, 
we can reduce the values of $m_c$, at the cost of a less efficient usage of the cache memories.
Figure~\ref{fig:gepm} reports the effect of this compromise, revealing that the pair
$m_c=(116,24)$ presents a better trade-off between balanced workload distribution and cache optimization.
For this operation, this concrete pair delivers 11.8--12.4 GFLOPS 
which is slightly above 80\% of the ideal peak performance.
A direct comparison of the top and
bottom traces in Figure~\ref{fig:traces} exposes the difference in workload distribution between
the executions with 
$m_c=(152,32)$ and
$m_c=(116,24)$, respectively. 

\subsection{Other BLIS-3 kernels with rectangular operands}

\begin{figure}[th!]
\begin{center}
\begin{tabular}{c}
\begin{minipage}[c]{\textwidth}
\includegraphics[width=0.5\textwidth]{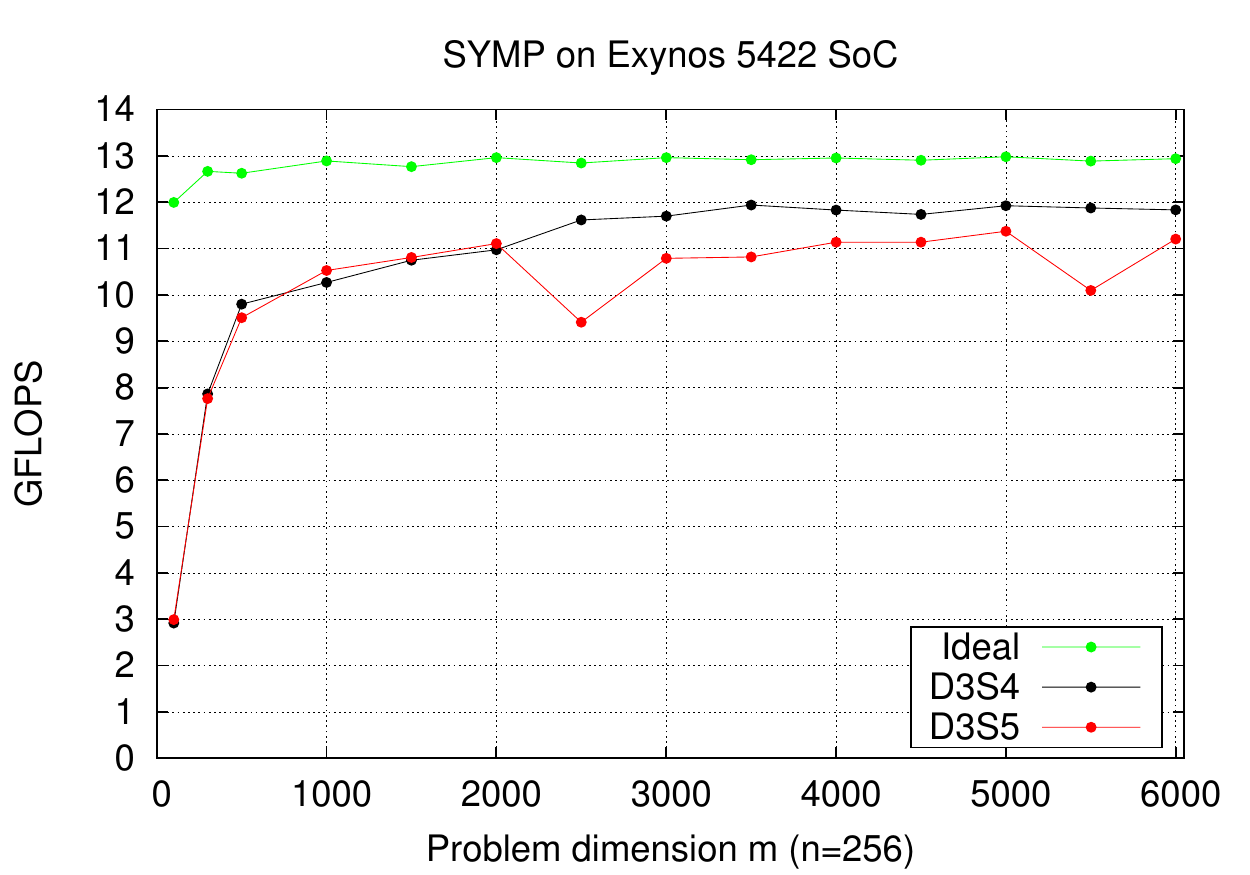}
\includegraphics[width=0.5\textwidth]{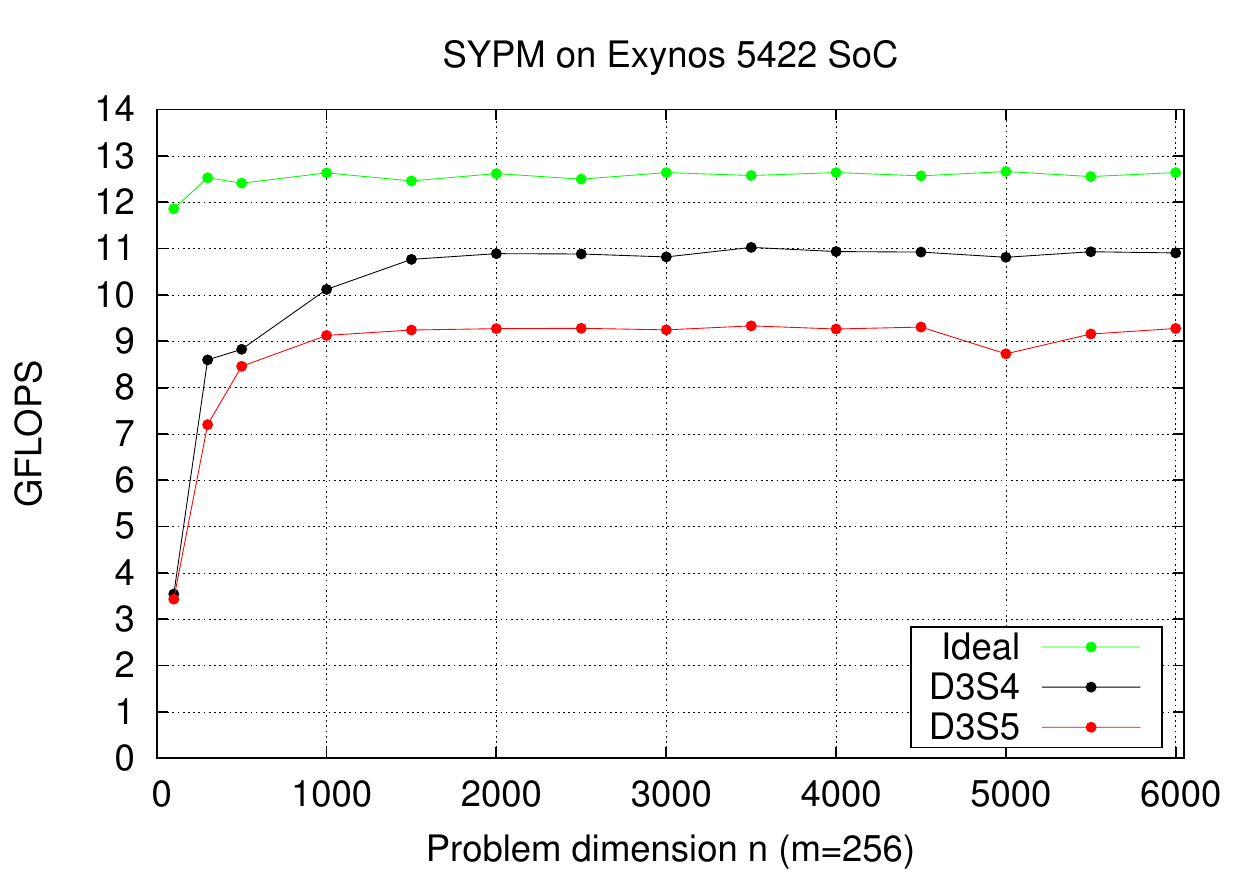}\\
\includegraphics[width=0.5\textwidth]{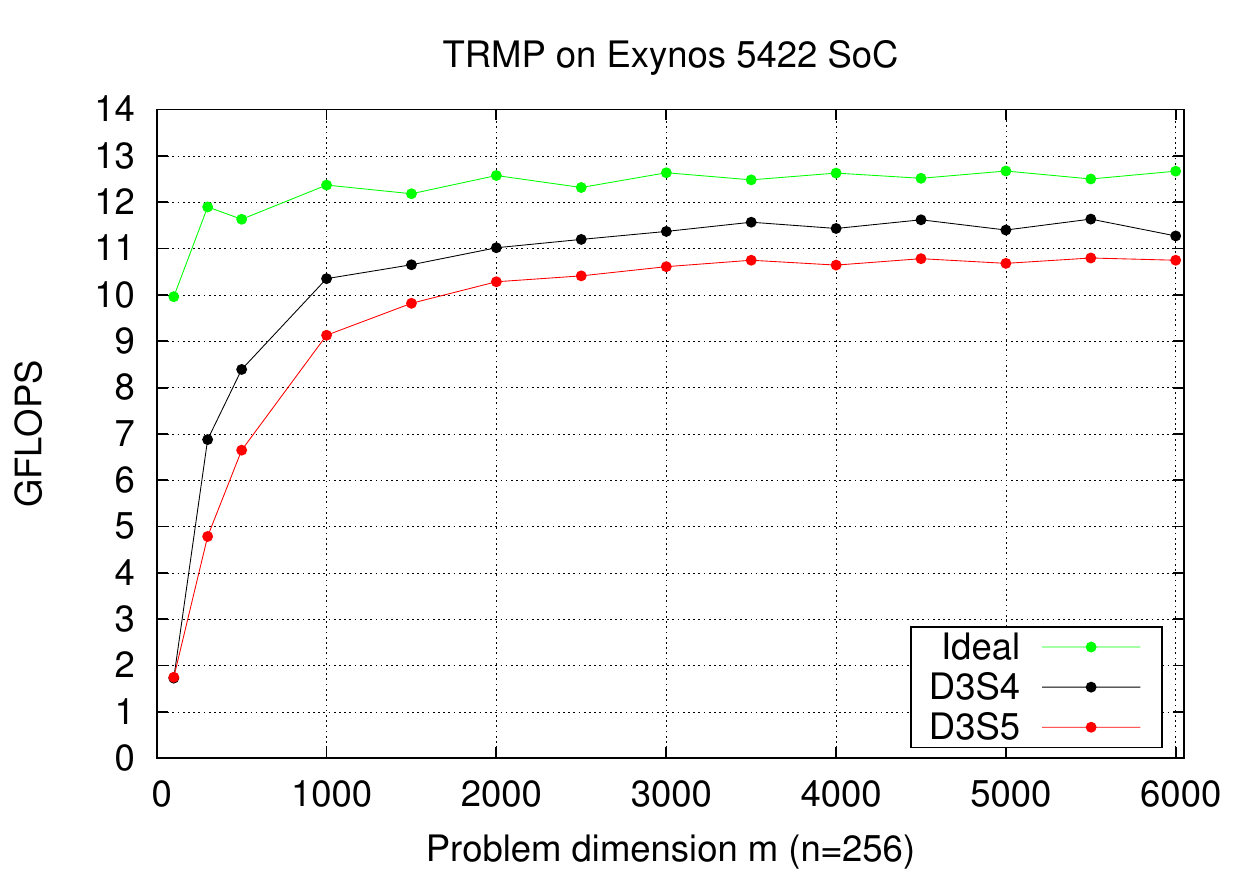}
\includegraphics[width=0.5\textwidth]{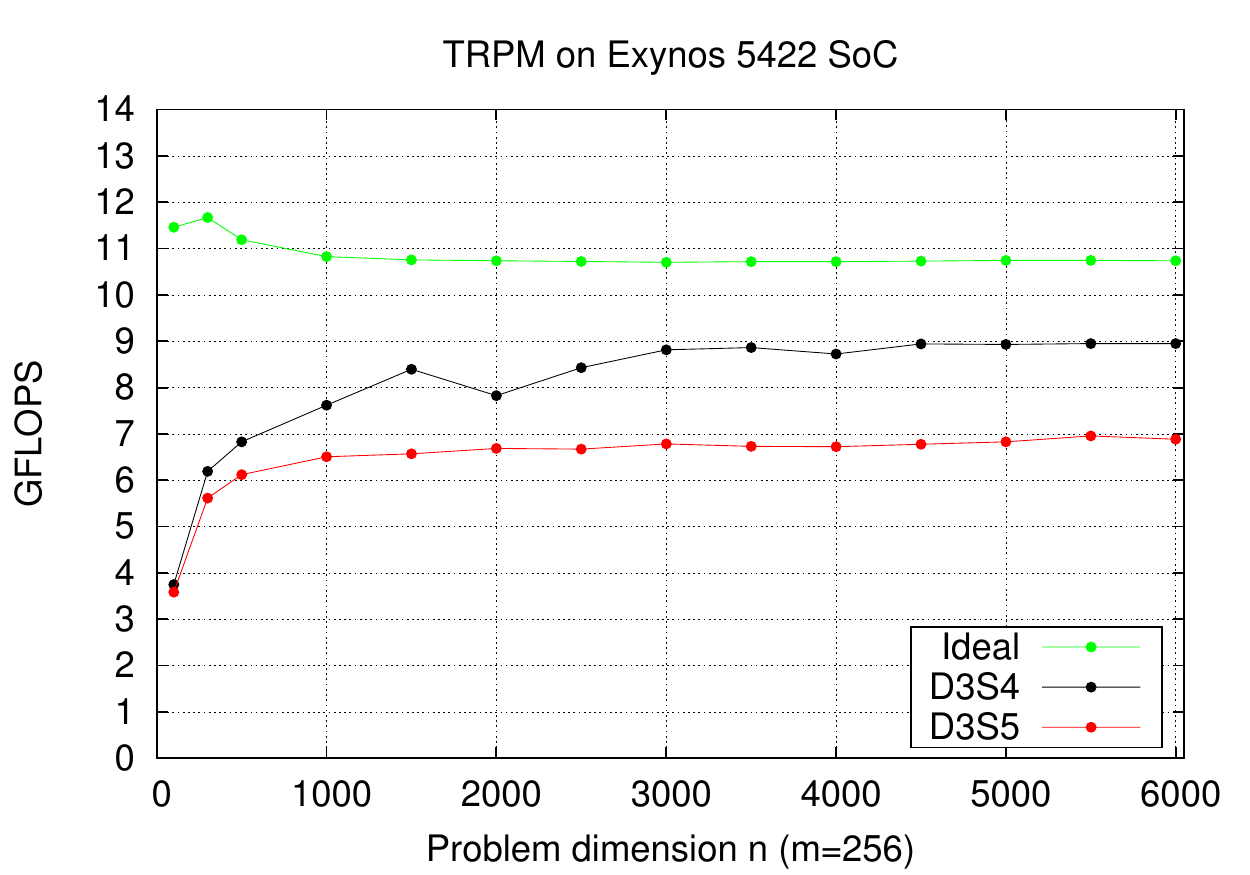}\\
\includegraphics[width=0.5\textwidth]{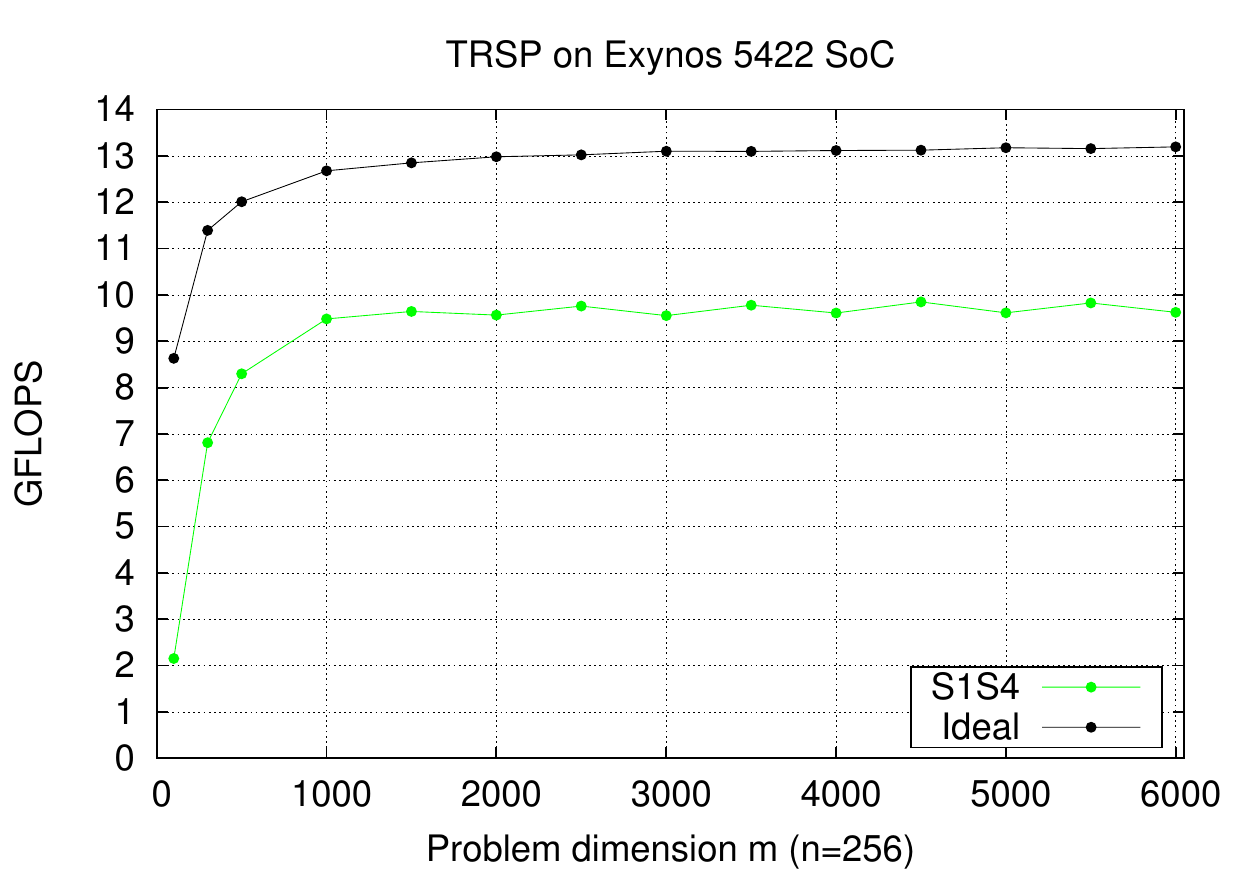}
\includegraphics[width=0.5\textwidth]{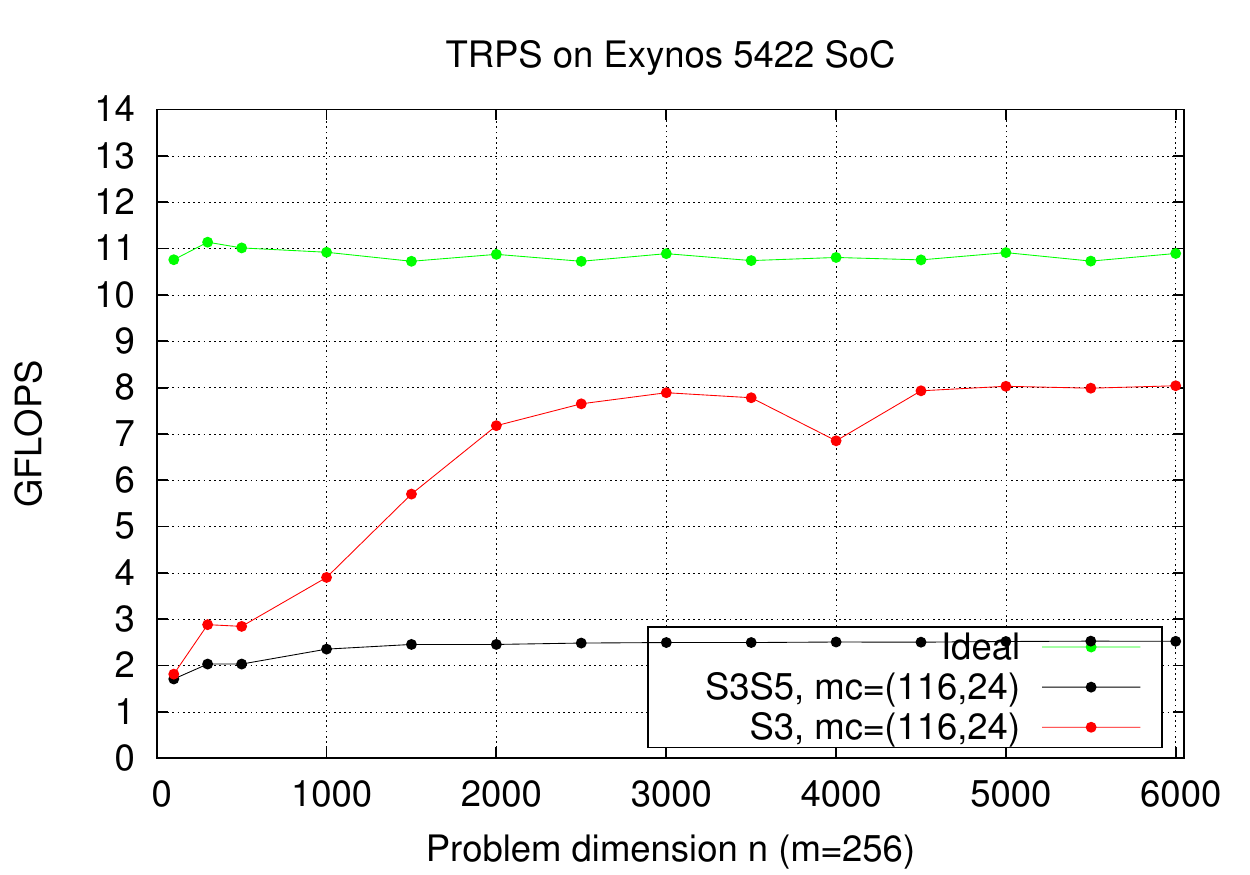}
\end{minipage}
\end{tabular}
\end{center}
\caption{Performance of two rectangular cases of 
         \symm (\symp for $C:=C+AB$ and \sypm for $C:=C+BA$),
         \trmm (\trmp for $B:=AB$ and \trpm for $B:=BA$), and
         \trsm (\trsp for $B:=A^{-1}B$ and \trsm for $B:=BA^{-1}$).}
\label{fig:sytr}
\end{figure}

Figure~\ref{fig:sytr} reports the performance of the BLIS kernels
for the symmetric matrix multiplication, the triangular matrix
multiplication, and the triangular system solve when applied to
two ``rectangular'' cases involving a symmetric/triangular matrix 
(see Table~\ref{tab:blis-3}): 
\begin{itemize}
\itemsep=0pt\parskip=0pt
\item 
      \symp, \trmp, \trsp 
when the symmetric/triangular matrix appears to the left-hand side of the operation (e.g., $C:=C+AB$ in \symp); 
\item 
      \sypm, \trpm, \trps 
when the symmetric/triangular matrix appears to the right-hand side of the operation (e.g., $C:=C+BA$ in \sypm).
\end{itemize}
The row and column 
dimensions of the symmetric/triangular matrix vary in the range ${\cal R}$ 
and the remaining problem size is fixed to $256$.
Therefore, when the matrix with special structure is to the right-hand side of the operator,
$m=256$. On the other hand, when this matrix is to the left-hand side, $n=256$. 
Also, in the left-hand side case, and for the same reasons argued for \gepm, 
we set $m_c=(116,24)$ for the ARM (Cortex-A15,Cortex-A7) cores.

Let us analyze the performance of the symmetric and triangular matrix multiplication kernels first.
From the plots in the top two rows of the figure, we can observe that {\tt D3S4} is still the best option for both
operations, independently of the side. 
When the problem dimension of the symmetric/triangular matrix
equals or exceeds 2000,
\symp delivers 11.0--11.9 GFLOPS,
\sypm 10.8--11.0 GFLOPS,
\trmp 11.0--11.6 GFLOPS, and
\trpm 7.8--8.9 GFLOPS.
Compared with the corresponding ideal peak performances, these values approximately represent fractions of
91\%, 
95\%,
90\%, and
80\%,
respectively.

The triangular system solve is a special case due to the dependencies intrinsic to this operation. For this particular
kernel, due to these dependencies,
the BLIS implementation cannot parallelize Loops~1 nor~4 when the triangular matrix is on the left-hand side.
For the same reasons, BLIS cannot parallelize Loops~3 nor~5 when this operator is on the right-hand side. 
Given these constraints, and the shapes of interest for the operands, 
we therefore select and evaluate the following
three simple {\em static} parallelization strategies. 
The first variant, {\tt S1S4}, 
is appropriate for \trsp and extracts coarse-grain parallelism from Loop~1 by statically dividing the complete
iteration space for this loop ($n$) between the two clusters, assigning $r_c=6\times$ more iterations to the ARM Cortex-A15 cluster
than to the slower ARM Cortex-A7 cluster. (This ratio~$r_c$ was experimentally identified in~\cite{asymBLIS} as a fair
representation of the performance difference between the two types of cores available in these clusters.)
In general, this strategy results in values for $n_c$ that are smaller than the theoretical optimal; however, given that
the Exynos 5422 SoC is not equipped with an L3 cache, the effect of this particular parameter is very small.
At a finer grain, this variant {\tt S1S4} statically distributes
the iteration space of Loop~4 among the cores within the same cluster.

The two other variants are designed for \trps, and they parallelize either Loop~3 only, or both Loops~3 and~5 
(denoted as {\tt S3} and {\tt S3S5}, respectively). In the first variant, the same ratio $r_c$ is applied 
to statically distribute the iterations of Loop~3 between the two types of cores. In the second variant,
the ratio statically partitions (coarse-grain parallelization)
the iteration space of the same loop between the two clusters and, internally (fine-grain parallelization), 
the workload comprised by Loop~5 is distributed among the cores of the same cluster.

The plots in the bottom row of Figure~\ref{fig:sytr} show that, for \trsp,
the parallelization of Loops 1+4 yields between 9.6 and 9.8~GFLOPS, which corresponds to about 74\% of the ideal peak performance;
for \trps, on the other hand, the parallelization of Loop~3 only is clearly superior to the combined parallelization
of Loops~3 and~5, offering 7.2--8.0~GFLOPS, which is within 65--75\% of the ideal peak performance.

\begin{figure}[th!]
\begin{center}
\begin{tabular}{c}
\begin{minipage}[c]{\textwidth}
\includegraphics[width=0.5\textwidth]{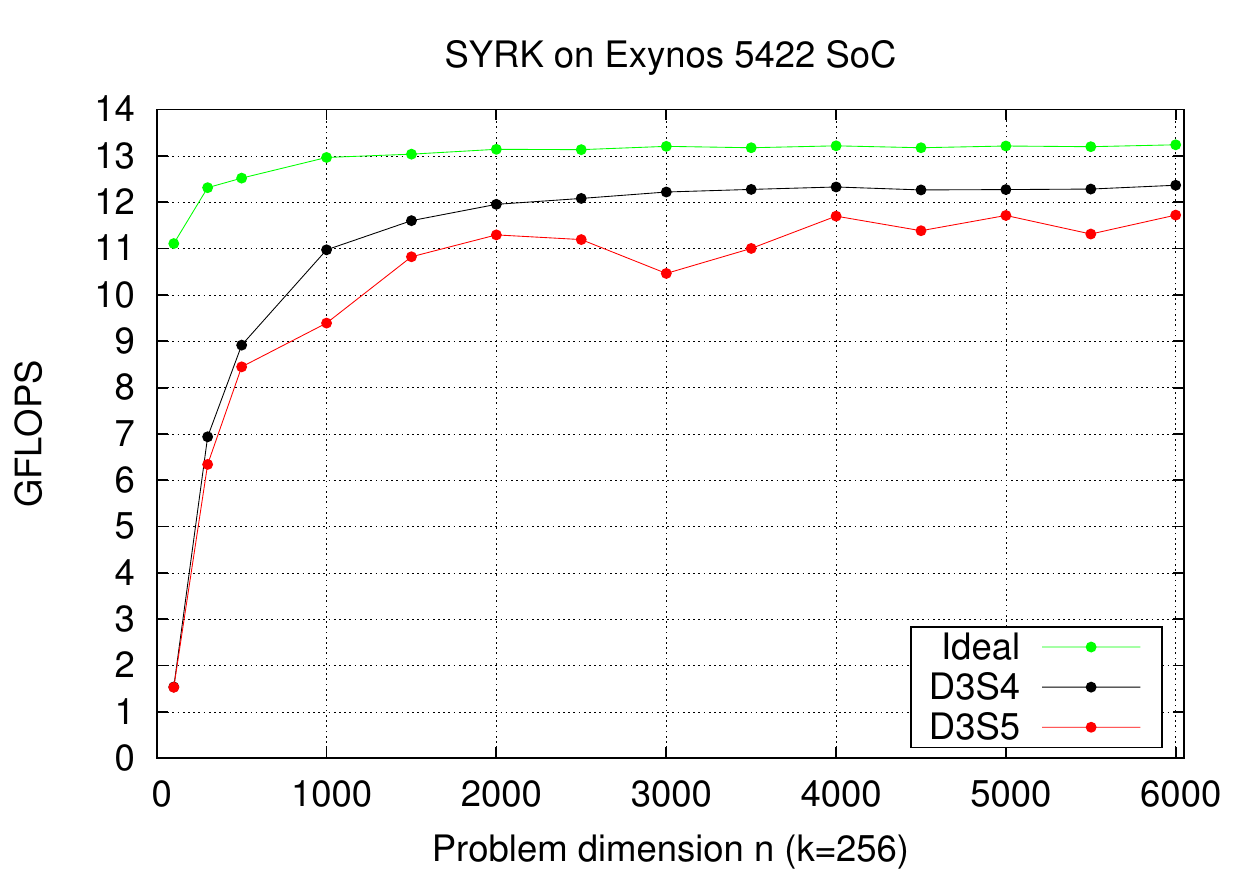}
\includegraphics[width=0.5\textwidth]{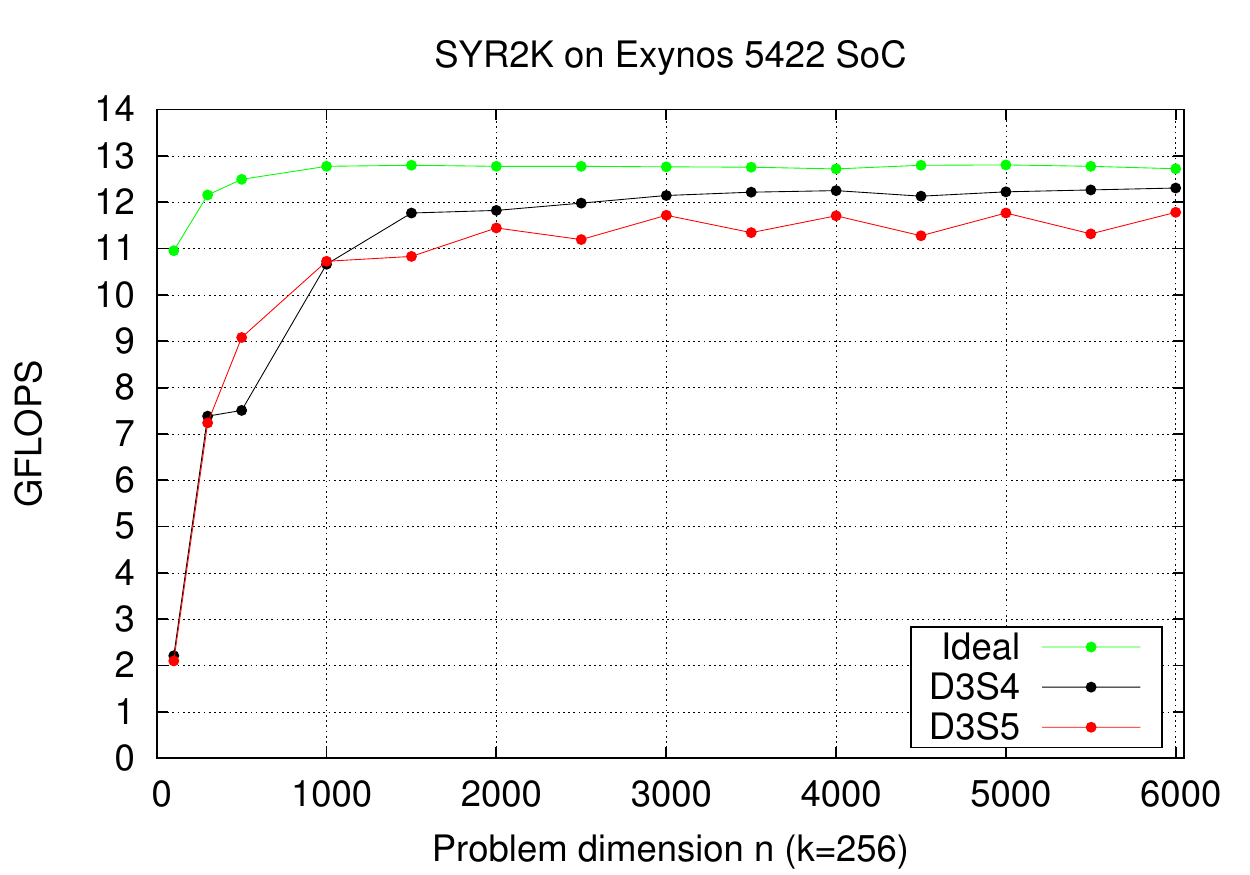}
\end{minipage}
\end{tabular}
\end{center}
\caption{Performance of a rectangular case of \syrk and \syrtk.}
\label{fig:syrk}
\end{figure}

To conclude the optimization and evaluation of the asymmetry-aware parallelization of BLIS,
Figure~\ref{fig:syrk}
illustrates the performance of the symmetric rank-$k$ and rank-$2k$ kernels,
when operating with rectangular operand(s) of dimension $n\times k$. 
For these two kernels, we vary $n$ in the range ${\cal R}$ and set $k=256$
(see Table~\ref{tab:blis-3}).
The results reveal high GFLOPS rates, similar to those observed for \gemm, 
and again slightly better for {\tt D3S4} compared with {\tt D3S5}.
In particular, the parallelization of Loops~3+4 renders
GFLOPS figures that are 12.0--12.4 and 11.8--12.3 for \syrk and \syrtk, respectively, when
$n$ is equal or larger than 2000. These performance rates are thus about 93\% of those
estimated for an the ideal scenario.

From a practical point of view,
the previous experimentation reveals {\tt D3S4} as the best
choice for all BLIS-3 kernels, except the triangular system solve; for the latter kernel we select
{\tt S1S4} when the operation/operands present a \trsp-shape or
{\tt S3} for operation/operands with \trps-shape. 
Additionally, in case $m$ is relatively small, our BLIS-3 kernels optimized for the Exynos 5422 SoC 
set $m_c=(116,24)$, but rely on the default $m_c=(152,32)$ otherwise.

\section{LAPACK for the Exynos 5422 SoC}
\label{sec:lapack}

Armed with the asymmetry-aware implementation of the BLIS-3 described in the previous section, 
we now target the execution of LAPACK on top of these optimized basic (level-3) kernels for the Exynos 5422 SoC.
For this purpose, we employ version~3.5.0 of LAPACK from netlib.
Here, our initial objective is to validate the soundness of our parallel version of BLIS-3
for the ARM big.LITTLE architecture, which was confirmed by successfully completing the correct execution of
the testing suite included in the LAPACK installation package.

In the following subsections,
we analyze the performance of our migration of LAPACK to the Exynos 5422 SoC. 
In this study, we are interested in assessing the performance of a ``plain''
migration; that is, one which does not carry out significant optimizations above the BLIS-3 layer.
We point out that this is the usual approach when there exist no native implementation of the LAPACK for
the target architecture, as is the case for the ARM big.LITTLE-based system.
The impact of limiting the optimizations to this layer will be exposed via three crucial 
dense linear algebra operations~\cite{GVL3}, illustrative of quite different outcomes: 
\begin{enumerate}
\itemsep=0pt\parskip=0pt
\item The Cholesky factorization for the solution of symmetric positive definite (s.p.d.) linear systems (routine \potrf).
\item The LU factorization (with partial row pivoting) for the solution of general linear systems (routine \getrf).
\item The reduction to tridiagonal form via similarity orthogonal transforms for the solution of symmetric eigenproblems
      (routine \sytrd).
\end{enumerate}
For brevity, we will only consider the real double precision case.

For the practical evaluation of these computational routines, 
we only introduced the following minor modifications in some of the LAPACK contents related with these routines:
\begin{enumerate}
\itemsep=0pt\parskip=0pt
\item We set the algorithmic block size {\tt NB} employed by these routines
      to $b={\tt NB}=256$ by adjusting the values returned by LAPACK routine {\sc ilaenv}.
\item For the Cholesky factorization, we modified the original LAPACK code 
      to obtain a right-looking variant of the algorithm~\cite{GVL3}, 
      numerically analogous to that implemented in the library, but which
      casts most of the flops in terms of a \syrk kernel with the shape and dimensions evaluated in the previous subsection,
      with $n$ in general larger than $k=b=256$.
\item For the Cholesky and LU factorizations, we changed the routines to (pseudo-)recursively 
      invoke the blocked variant of the code (with
      block size $\tilde{b}=32$) in order to process
      the ``current'' diagonal block and column panel, respectively~\cite{GVL3}.
\end{enumerate}

\subsection{Cholesky factorization}

\begin{figure}[th!]
\begin{center}
\includegraphics[width=0.6\textwidth]{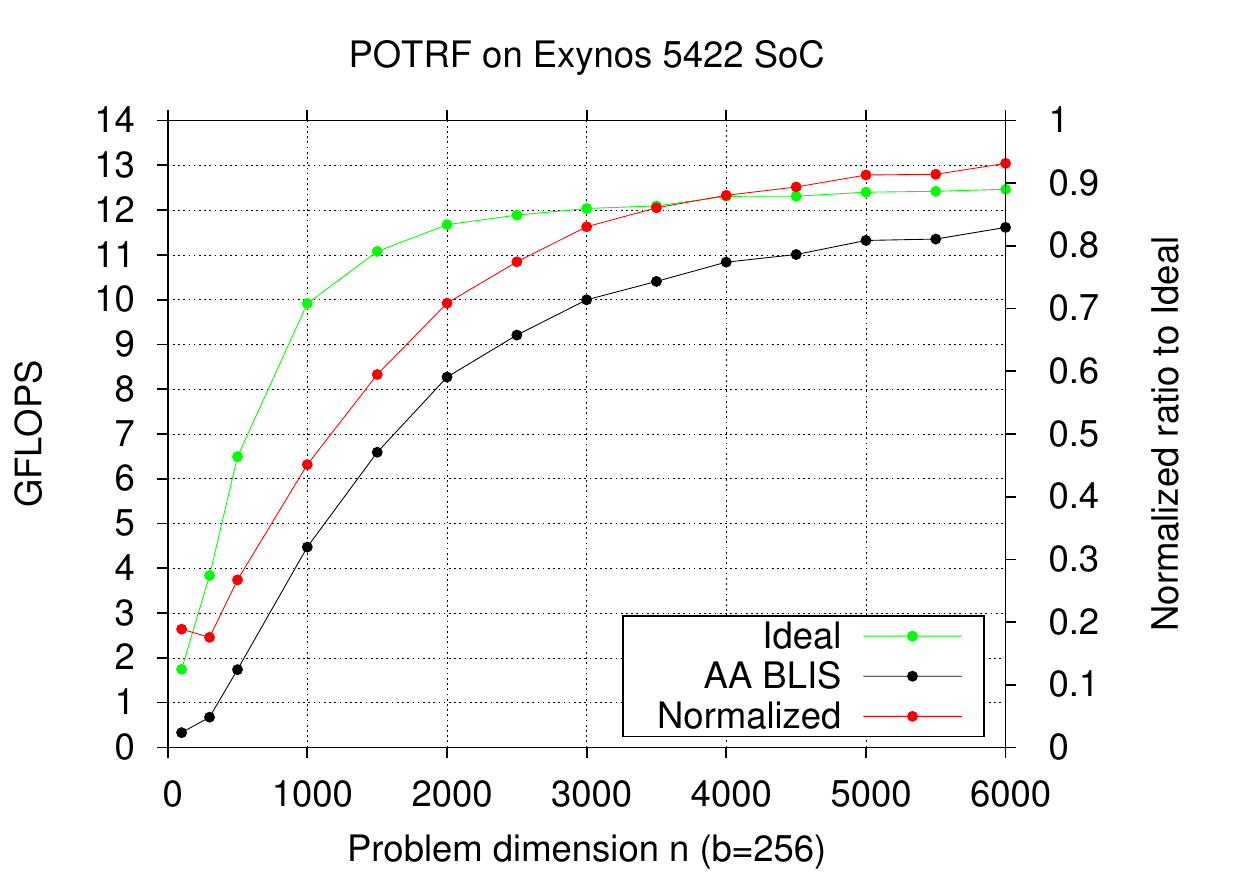}
\end{center}
\caption{Performance of \potrf for the solution s.p.d. linear systems.}
\label{fig:chol}
\end{figure}

Figure~\ref{fig:chol} reports the GFLOPS rates obtained with (our right-looking variant of 
the routine for) the Cholesky factorization (\potrf), 
executed on top of the asymmetry-aware BLIS-3 ({\tt AA BLIS}), 
when applied to compute the upper triangular Cholesky factors of matrices of dimension $n$ in the range ${\cal R}$. 
Following the kind of comparison done for the BLIS-3,
in the plot we also include the performance estimated for the ideal configuration (scale in the left-hand side
$y$-axis). Furthermore, we offer the 
ratio that the actual GFLOPS rate represents compared with that estimated under the ideal conditions
(line labeled as {\tt Normalized}, with scale in the right-hand side $y$-axis).

For this particular factorization, as the problem dimension grows, the gap between the ideal peak performance
and the actual GFLOPS rate rapidly shrinks. 
This is quantified in the columns labeled as {\tt Normalized} in Table~\ref{tab:ls}, which reflect
the numerical values represented by the normalized curve in Figure~\ref{fig:chol}. Here, for example, 
the implementation obtains over 70\% and 88\% of the ideal peak performance for
$n=2000$ and $n=3000$, respectively.

This appealing behaviour is well explained by considering how this algorithm, rich in BLAS-3 kernels, proceeds. 
Concretely, at each iteration, the right-looking version decomposes the calculation into three kernels, 
with one of them being
a symmetric rank-$k$ update (\syrk) involving a row panel of $k=b$ rows~\cite{GVL3}.
Furthermore, as $n$ grows, the cost of this update rapidly dominates the total cost of the decomposition;
see the columns for the normalized flops in Table~\ref{tab:ls}.
As a result,
the performance of this variant of the Cholesky factorization approaches that of \syrk; see
in Figure~\ref{fig:syrk}. 
Indeed, it is quite remarkable that, for $n=6000$, the implementation of the Cholesky factorization
attains slightly more than 93\% of the ideal peak performance, which is basically the same
fraction of the ideal peak observed for \syrk and a problem of dimension $n=6000, k=256$.

\begin{table}[th!]
\begin{center}
{\footnotesize
\begin{tabular}{|c||c|c||c|c|}
\hline 
        & \multicolumn{2}{c||}{\potrf}  & \multicolumn{2}{c|}{\getrf}         \\
$n$     & Normalized       & Normalized     & Normalized       & Normalized  \\ 
        & GFLOPS           & flops of \syrk & GFLOPS           & flops of \gepp \\ \hline \hline
   ~100 &  18.86 & ~0.00 & 14.08 & ~0.00 \\ 
   ~300 &  17.55 & ~5.50 & ~4.90 & ~5.50 \\ 
   ~500 &  26.73 & 36.67 & 12.56 & 36.67 \\ 
   1000 &  45.12 & 64.97 & 28.59 & 64.97 \\ 
   1500 &  59.49 & 75.90 & 45.22 & 75.90 \\ \hline
   2000 &  70.85 & 81.65 & 53.27 & 81.65 \\ 
   2500 &  77.46 & 85.18 & 60.70 & 85.18 \\ 
   3000 &  83.06 & 87.58 & 65.11 & 87.58 \\ 
   3500 &  86.05 & 89.31 & 69.36 & 89.31 \\ 
   4000 &  88.06 & 90.61 & 70.80 & 90.61 \\ \hline
   4500 &  89.39 & 91.63 & 74.51 & 91.63 \\ 
   5000 &  91.29 & 92.46 & 75.21 & 92.46 \\ 
   5500 &  91.42 & 93.15 & 80.00 & 93.15 \\ 
   6000 &  93.16 & 93.69 & 84.30 & 93.69 \\ \hline
\end{tabular}
}
\end{center}
\caption{Performance of matrix factorizations for the solution of s.p.d. and general linear systems
         (\potrf and \getrf, respectively) normalized with respect to the ideal peak performance (in \%); 
         and corresponding theoretical costs of the underlying basic building blocks 
         \syrk and \gepp normalized with respect to the total factorization cost (in \%).}
\label{tab:ls}
\end{table}

\subsection{LU factorization}

\begin{figure}[th!]
\begin{center}
\includegraphics[width=0.6\textwidth]{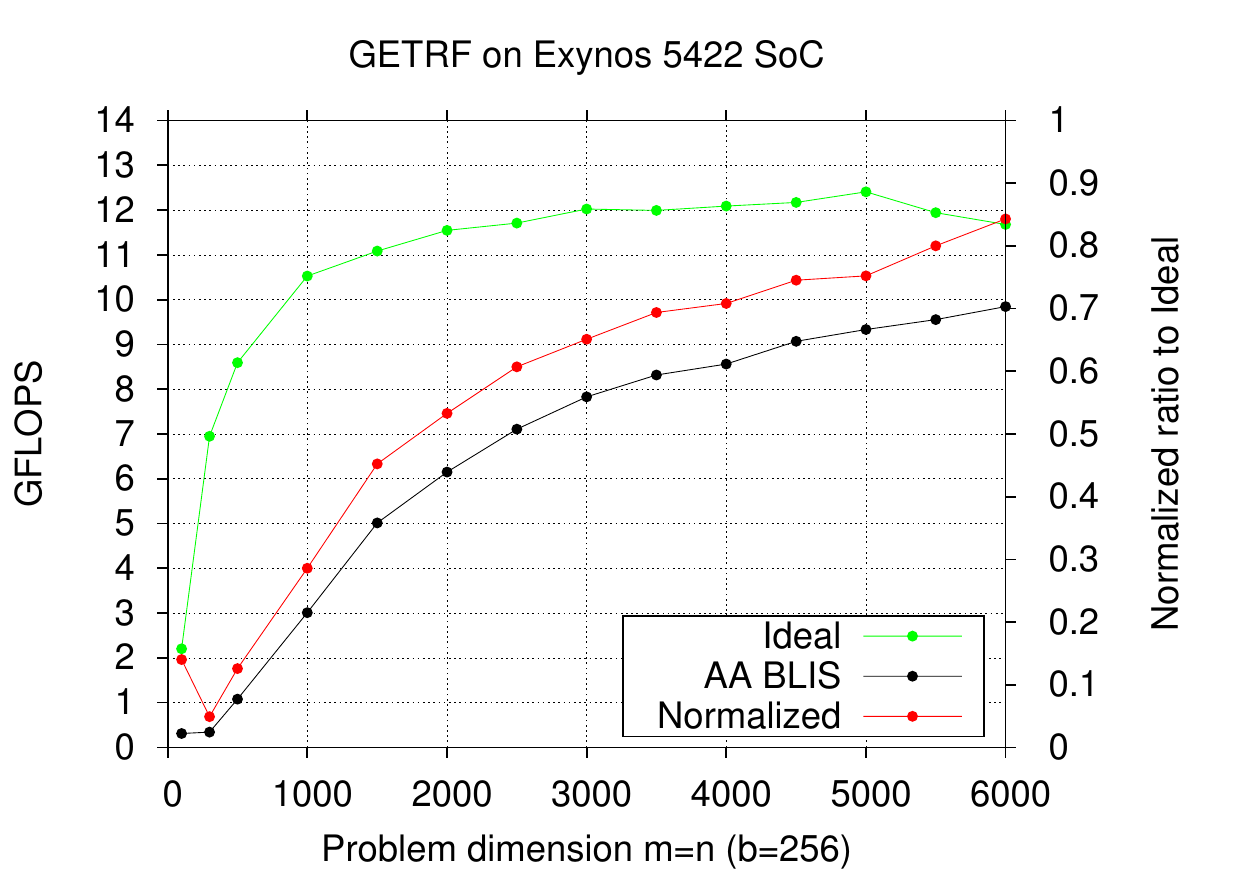}
\end{center}
\caption{Performance of \getrf for the solution general linear systems.}
\label{fig:lu}
\end{figure}

Figure~\ref{fig:lu} displays the GFLOPS attained by the
routine for the LU factorization with partial row pivoting (\getrf),  linked with
the asymmetry-aware BLIS-3,
when applied to decompose square matrices of dimension $m=n$ in~${\cal R}$. 

The actual performance of the LU factorization follows the same general trend observed for the Cholesky factorization,
though there are some differences worth of being justified. 
First, the migration of the Cholesky factorization to the Exynos 5422 SoC 
was a story of success, while the LU factorization reflects a less pleasant case.
For example, the routine for the LU factorization attains over 53\% and 65.11\% of the ideal peak performance
for $n=2000$ and $n=3000$, respectively. 
Compared with this, the Cholesky factorization attained more than 70\% and 83\% at the same points.
A case-by-case comparison can be quickly performed by inspecting the columns reporting the normalized GFLOPS
for each factorization in Table~\ref{tab:ls}.

Let us discuss this further. Like \potrf, routine \getrf casts most flops in terms of efficient BLAS-3 kernels,
in this case the panel-panel multiplication \gepp.
Nonetheless, its moderate performance behavior 
lies in the high practical cost (i.e., execution time) of the column panel factorization that is present at each iteration
of the LU procedure. In particular, this panel factorization stands in the critical path of the algorithm
and exhibits a limited amount of concurrency, 
easily becoming a serious bottleneck
when the number of cores is large relative to the problem dimension. 
To illustrate this point, the LU factorization of the panel
takes 27.79\% of the total time during a parallel factorization of a matrix of order $m=n=3000$.
Compared with this, the decomposition of the diagonal block present in the Cholesky factorization, which plays an analogous
role, represents only 10.42\% of the execution time for the same problem dimension.

This is a known problem for which there exist {\em look-ahead} variants of the factorization procedure that overlap
the update of the trailing submatrix with the factorization of the next panel,
thus eliminating the 
latter from the critical path~\cite{Str98}. However, introducing a static look-ahead strategy into the code
is by no means straight-forward, and therefore
is in conflict with our goal of assessing the efficiency of a plain migration of LAPACK.
As an alternative, one could rely on a runtime to produce the same effect, by (semi-)automatically
introducing a sort of dynamic look-ahead
into the execution of the factorization. However, the application of a runtime to a legacy code is not
as simple as it may sound and, as argued in the discussion of related work, the development of asymmetry-aware runtimes is
still immature.

\subsection{Reduction to tridiagonal form}

\begin{figure}[th!]
\begin{center}
\includegraphics[width=0.6\textwidth]{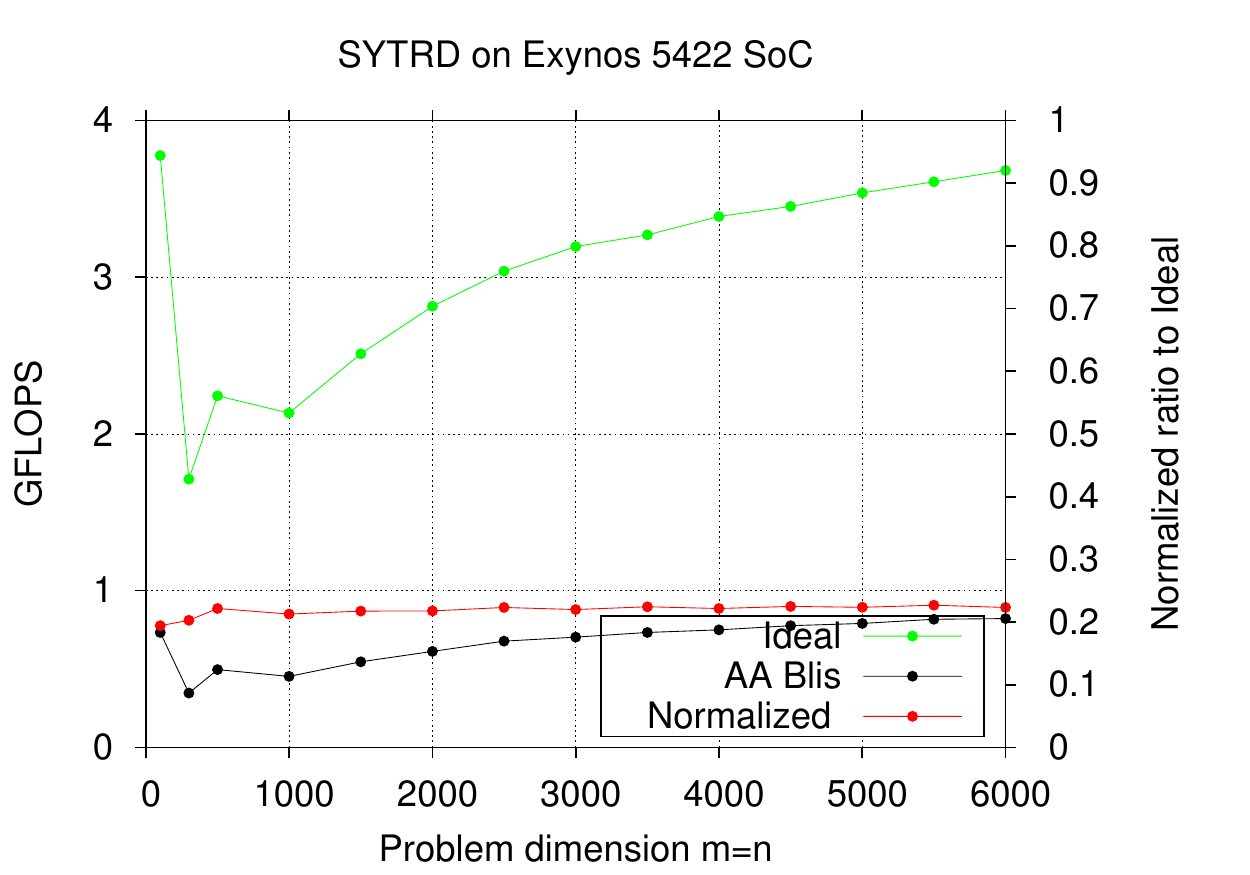}
\end{center}
\caption{Performance of \sytrd for the reduction to tridiagonal form.}
\label{fig:sytrd}
\end{figure}

To conclude this section, Figure~\ref{fig:sytrd} reports the performance behaviour of the LAPACK routine
for the reduction to tridiagonal form, \sytrd. 
Here, we also execute the routine on top of the asymmetry-aware BLIS-3; and apply it to (the upper triangle of) 
symmetric matrices of dimension $n$ in ${\cal R}$.

The first difference to discuss between the results observed for this routine and those of the Cholesky and
LU factorization is the scale of the left-hand side $y$-axis, with an upper limit at 4 GFLOPS for \sytrd against
14 for the other two. The reason is that the reduction procedure underlying \sytrd  casts half of its 
flops in terms of the symmetric matrix-vector product, \symv, a memory-bound kernel that belongs to the BLAS-2.
Concretely, this kernel roughly performs~4 flops only per memory access 
and cannot take full advantage of the FPUs available in the system, which
will be stalled most of the time waiting for data (the symmetric matrix entries) from memory.
A second aspect to point out is the low fraction of the ideal peak performance attained with the asymmetry-aware implementation.
Unfortunately, even though \sytrd performs the remaining 50\% of its flops via the highly efficient \syrtk, 
the execution time of this other half is practically negligible compared with the execution of the
\symv kernels (for the problem dimensions evaluated in the paper, less than 5\%).
In addition, we note that BLIS does not provide parallel versions of the \symv (nor any other routine from the 
Level-1 and Level-2 BLAS), which helps to explain the low performance attained with our plain migration of this 
LAPACK routine on top of a parallel BLIS-3 implementation.

\section{Conclusions and Future Work}
\label{sec:remarks}

We have leveraged the flexibility of the BLIS framework in order to 
introduce an asymmetry-aware (and in most cases) high performance implementation of the BLAS-3 for 
AMPs, such as the ARM big.LITTLE SoC, that takes into consideration the operands' dimensions and shape.
The key to our development is the integration of a coarse-grain scheduling policy, which
dynamically distributes the workload between the two core types present in this architecture, combined with a
complementary static schedule that repartitions this work among the cores of the same type.
Our experimental results on the target platform in general show considerable performance acceleration for
the BLAS-3 kernels, and more moderate for the triangular system solve. 

In addition, we have migrated a legacy implementation of LAPACK that leverages the asymmetry-aware
BLIS-3 to run on the target AMP. In doing so, we have explored the benefits and drawbacks of conducting a
simple (plain) migration which does not perform any major optimizations in LAPACK.
Our experimentation with three major routines from LAPACK illustrates three distinct scenarios (cases), 
ranging from a compute-bound operation/routine (Cholesky factorization) 
where high performance is easily attained from this plain migration;
to a compute-bound operation (LU factorization) where the same level of success will require a significant 
reorganization of the code that introduces an advanced scheduling mechanism; and, finally, 
a memory-bound case (reduction to tridiagonal form)
where an efficient parallelization of the BLAS-2 is key to obtain even moderate performance.

As part of future work, 
we will explore alternative parallelization strategies that better suite the triangular system solve kernel;
we plan to introduce an asymmetry-aware static look-ahead scheduling into one-sided panel-operations such as the LU and
QR factorizations; and we will develop an asymmetry-conscious version of the BLAS-2 from the BLIS framework.

\section*{Acknowledgments}

The researchers from Universidad Jaume~I 
were supported by projects CICYT TIN2011-23283 and TIN2014-53495-R of
MINECO and FEDER, and the FPU program of MECD.
The researcher from Universidad Complutense de Madrid 
was supported by project CICYT TIN2012-32180.
The researcher from Universitat Polit\`ecnica de Catalunya was supported
by projects  TIN2012-34557 from  the Spanish  Ministry of  Education and
2014 SGR 1051  from the Generalitat de  Catalunya, Dep. d\'~Innovaci\'o,
Universitats i Empresa.

  \bibliographystyle{elsarticle-num} 
  \bibliography{enrique,energy,asymmetric}

\begin{thebibliography}{10}
\expandafter\ifx\csname url\endcsname\relax
  \def\url#1{\texttt{#1}}\fi
\expandafter\ifx\csname urlprefix\endcsname\relax\def\urlprefix{URL }\fi
\expandafter\ifx\csname href\endcsname\relax
  \def\href#1#2{#2} \def\path#1{#1}\fi

\bibitem{dem97}
J.~Demmel, Applied Numerical Linear Algebra, Society for Industrial and Applied
  Mathematics, 1997.

\bibitem{lapack}
E.~\mbox{Anderson et al}, {LAPACK} Users' guide, 3rd Edition, SIAM, 1999.

\bibitem{blas1}
C.~L. Lawson, R.~J. Hanson, D.~R. Kincaid, F.~T. Krogh, Basic linear algebra
  subprograms for {F}ortran usage, ACM Transactions on Mathematical Software
  5~(3) (1979) 308--323.

\bibitem{blas2}
J.~J. Dongarra, J.~Du~Croz, S.~Hammarling, R.~J. Hanson, An extended set of
  {FORTRAN} basic linear algebra subprograms, ACM Transactions on Mathematical
  Software 14~(1) (1988) 1--17.

\bibitem{blas3}
J.~J. Dongarra, J.~Du~Croz, S.~Hammarling, I.~Duff, A set of level 3 basic
  linear algebra subprograms, ACM Trans. Math. Softw. 16~(1) (1990) 1--17.

\bibitem{acml}
AMD, {AMD} {C}ore {M}ath {L}ibrary,
  \url{http://developer.amd.com/tools/cpu/acml/pages/default.aspx} (2012).

\bibitem{essl}
IBM, {E}ngineering and {S}cientific {S}ubroutine {L}ibrary,
  \url{http://www.ibm.com/systems/software/essl/} (2012).

\bibitem{mkl}
{Intel~Corp.}, Intel math kernel library ({MKL}) 11.0,
  \url{http://software.intel.com/en-us/intel-mkl} (2014).

\bibitem{cublas}
NVIDIA, {CUDA} basic linear algebra subprograms,
  \url{https://developer.nvidia.com/cuBLAS} (2014).

\bibitem{atlas}
R.~C. Whaley, J.~J. Dongarra, Automatically tuned linear algebra software, in:
  Proceedings of SC'98, 1998.

\bibitem{Goto:2008:AHP}
K.~Goto, R.~van~de Geijn, Anatomy of a high-performance matrix multiplication,
  ACM Trans. Math. Softw. 34~(3) (2008) 12:1--12:25.

\bibitem{OpenBLAS}
Open{BLAS}, \url{http://xianyi.github.com/OpenBLAS/} (2012).

\bibitem{BLIS1}
F.~G. {V}an {Z}ee, R.~A. {v}an~{d}e {G}eijn, {BLIS}: A framework for generating
  blas-like libraries, {ACM} Trans. Math. Soft.To appear.

\bibitem{Den74}
R.~Dennard, F.~Gaensslen, V.~Rideout, E.~Bassous, A.~LeBlanc, Design of
  ion-implanted {MOSFET}'s with very small physical dimensions, Solid-State
  Circuits, IEEE Journal of 9~(5) (1974) 256--268.

\bibitem{Moo65}
G.~Moore, Cramming more components onto integrated circuits, Electronics 38~(8)
  (1965) 114--117.

\bibitem{Kum04}
R.~Kumar, D.~M. Tullsen, P.~Ranganathan, N.~P. Jouppi, K.~I. Farkas,
  Single-{ISA} heterogeneous multi-core architectures for multithreaded
  workload performance, in: Proc. 31st Annual Int. Symp. on Computer
  Architecture, ISCA'04, 2004, p.~64.

\bibitem{Hil08}
M.~Hill, M.~Marty, Amdahl's law in the multicore era, Computer 41~(7) (2008)
  33--38.

\bibitem{Mor06}
T.~Morad, U.~Weiser, A.~Kolodny, M.~Valero, E.~Ayguade, Performance, power
  efficiency and scalability of asymmetric cluster chip multiprocessors,
  Computer Architecture Letters 5~(1) (2006) 14--17.

\bibitem{Win10}
J.~A. Winter, D.~H. Albonesi, C.~A. Shoemaker, Scalable thread scheduling and
  global power management for heterogeneous many-core architectures, in: Proc.
  19th Int. Conf. Parallel Architectures and Compilation Techniques, PACT'10,
  2010, pp. 29--40.

\bibitem{asymBLIS}
S.~Catal{\'{a}}n, F.~D. Igual, R.~Mayo, R.~Rodr{\'{\i}}guez{-}S{\'{a}}nchez,
  E.~S. Quintana{-}Ort{\'{\i}},
  \href{http://arxiv.org/abs/1506.08988}{Architecture-aware configuration and
  scheduling of matrix multiplication on asymmetric multicore processors},
  ArXiv e-prints 1506.08988.
\newline\urlprefix\url{http://arxiv.org/abs/1506.08988}

\bibitem{IntelQuickIA}
N.~Chitlur, G.~Srinivasa, S.~Hahn, P.~Gupta, D.~Reddy, D.~Koufaty, P.~Brett,
  A.~Prabhakaran, L.~Zhao, N.~Ijih, S.~Subhaschandra, S.~Grover, X.~Jiang,
  R.~Iyer, Quickia: Exploring heterogeneous architectures on real prototypes,
  in: High Performance Computer Architecture (HPCA), 2012 IEEE 18th
  International Symposium on, 2012, pp. 1--8.
\newblock \href {http://dx.doi.org/10.1109/HPCA.2012.6169046}
  {\path{doi:10.1109/HPCA.2012.6169046}}.

\bibitem{BLIS2}
F.~G. Van~Zee, T.~M. Smith, B.~Marker, T.~M. Low, R.~A. van~de Geijn, F.~D.
  Igual, M.~Smelyanskiy, X.~Zhang, M.~Kistler, V.~Austel, J.~Gunnels,
  L.~Killough, The {BLIS} framework: Experiments in portability, {ACM} Trans.
  Math. Soft.~To appear. Available at
  \url{http://www.cs.utexas.edu/users/flame}.

\bibitem{BLIS3}
T.~M. Smith, R.~van~de Geijn, M.~Smelyanskiy, J.~R. Hammond, F.~G. Van~Zee,
  Anatomy of high-performance many-threaded matrix multiplication, in: Proc.
  IEEE 28th Int. Parallel and Distributed Processing Symp., IPDPS'14, 2014, pp.
  1049--1059.

\bibitem{cilkweb}
{Cilk} project home page, \url{http://supertech.csail.mit.edu/cilk/}.

\bibitem{ompssweb}
{OmpSs} project home page, \url{http://pm.bsc.es/ompss}, last visit: July 2015.

\bibitem{starpuweb}
{StarPU} project home page, \url{http://runtime.bordeaux.inria.fr/StarPU/}.

\bibitem{plasmaweb}
{PLASMA} project home page, \url{http://icl.cs.utk.edu/plasma/}.

\bibitem{magmaweb}
{MAGMA} project home page, \url{http://icl.cs.utk.edu/magma/}.

\bibitem{kaapiweb}
{Kaapi} project home page, \url{https://gforge.inria.fr/projects/kaapi}, last
  visit: July 2015.

\bibitem{flameweb}
{FLAME} project home page, \url{http://www.cs.utexas.edu/users/flame/}.

\bibitem{OmpSsbigLITTLE}
K.~Chronaki, A.~Rico, R.~M. Badia, E.~Ayguad{\'e}, J.~Labarta, M.~Valero,
  Criticality-aware dynamic task scheduling for heterogeneous architectures,
  in: Proceedings of ICS'15, 2015.

\bibitem{asymChol}
L.~Costero, F.~D. Igual, K.~Olcoz, E.~S. Quintana-Ort\'i,
  \href{http://arxiv.org/abs/1590.02058}{Exploiting asymmetry in arm big.little
  architectures for dense linear algebra using conventional task schedulers},
  ArXiv e-prints 1590.02058.
\newline\urlprefix\url{http://arxiv.org/abs/1590.02058}

\bibitem{BadiaHLPQQ09}
R.~M. Badia, J.~R. Herrero, J.~Labarta, J.~M. P\'erez, E.~S.
  Quintana-Ort\'{\i}, G.~Quintana-Ort\'{\i}, Parallelizing dense and banded
  linear algebra libraries using {SMPSs}, Conc. and Comp.: Pract. and Exper. 21
  (2009) 2438--2456.

\bibitem{Buttari200938}
A.~Buttari, J.~Langou, J.~Kurzak, , J.~Dongarra, A class of parallel tiled
  linear algebra algorithms for multicore architectures, Parallel Computing
  35~(1) (2009) 38--53.

\bibitem{Quintana:2008:PMA}
G.~Quintana-Ort{\'\i}, E.~S. Quintana-Ort{\'\i}, R.~A. van~de Geijn, F.~G.
  Van~Zee, E.~Chan, Programming matrix algorithms-by-blocks for thread-level
  parallelism, ACM Trans. Math. Softw. 36~(3) (2009) 14:1--14:26.

\bibitem{extrae}
H.~Servat, G.~Llort, {Extrae} user guide manual for version 3.2.1,
  \url{http://www.bsc.es/computer-sciences/extrae} (2015).

\bibitem{GVL3}
G.~H. Golub, C.~F.~V. Loan, Matrix Computations, 3rd Edition, The Johns Hopkins
  University Press, Baltimore, 1996.

\bibitem{Str98}
P.~Strazdins, A comparison of lookahead and algorithmic blocking techniques for
  parallel matrix factorization, Tech. Rep. TR-CS-98-07, Department of Computer
  Science, The Australian National University, {Canberra} 0200 {ACT},
  {Australia} (1998).

\end{thebibliography}


%
%
%
\end{document}